\documentclass[aps,prb,superscriptaddress,showpacs,twocolumn]{revtex4-1}

\usepackage{graphicx}
\usepackage{amsmath}
\usepackage{bbm}
\usepackage{xspace}
\usepackage{color}
\usepackage{sidecap}
\usepackage{hyperref}

\usepackage[normalem]{ulem} 

\newcommand{\ie}{i.e.\@\xspace}
\newcommand{\eg}{e.g.\@\xspace}

\newcommand{\ve}[1]{{\bf #1}}

\newcommand{\nag}{{\phantom{\dagger}}}

\newcommand{\eqw}[1]{(\ref{#1})}
\newcommand{\eq}[1]{Eq.\thinspace{}(\ref{#1})}
\newcommand{\eqq}[2]{Eqs.\thinspace{}(\ref{#1}) and (\ref{#2})}

\newcommand{\den}{\rho}

\newcommand{\tab}[1]{Tab.\thinspace{}\ref{#1}}

\newcommand{\fig}[1]{Fig.\thinspace{}\ref{#1}}

\newcommand{\fc}[1]{({#1})}
\newcommand{\figc}[2]{Fig.\thinspace{}\ref{#1}\thinspace{}\fc{#2}}
\newcommand{\figcc}[3]{Fig.\thinspace{}\ref{#1}\thinspace{}\fc{#2} and \fc{#3}}

\begin{document}

\title{Clustered Wigner crystal phases of cold polar molecules in arrays of one-dimensional tubes}

\author{Michael Knap}
\email[]{michael.knap@tugraz.at}
\affiliation{Institute of Theoretical and Computational Physics, Graz University of Technology, 8010 Graz, Austria}
\affiliation{Department of Physics, Harvard University, Cambridge MA 02138, USA}

\author{Erez Berg}
\affiliation{Department of Physics, Harvard University, Cambridge MA 02138, USA}

\author{Martin Ganahl}
\affiliation{Institute of Theoretical and Computational Physics, Graz University of Technology, 8010 Graz, Austria}

\author{Eugene Demler}
\affiliation{Department of Physics, Harvard University, Cambridge MA 02138, USA}

\date{\today}

\begin{abstract}
We analyze theoretically polar molecules confined in planar arrays
of one dimensional tubes. In the classical limit, if the number of
tubes is finite, new types of ``clustered Wigner crystals'' with
increasingly many molecules per unit cell can be stabilized by
tuning the in-plane angle between the dipolar moments and the tube
direction. Quantum mechanically, these phases melt into distinct
``clustered Luttinger liquids.'' We calculate the phase diagram of the
system and study the quantum melting of the clustered phases. We
find that the requirements for exploring these phases are
reachable in current experiments and discuss possible experimental
signatures.
\end{abstract}

\pacs{
05.30.-d, 
03.75.Hh, 
67.85.-d, 
34.20.-b 
}

\maketitle

\section{Introduction}

Systems with competing long-range interactions often exhibit
structures with emergent large length scales. Some examples include
the formation of bubble and stripe domains in Langmuir-Blodgett
films or in thin
ferromagnetic layers,~\cite{ng_stability_1995,marchenko_1986} and
the chain formation of magnetic particles in three-dimensional
ferrofluids.~\cite{rosenweig_ferrohydrodynamics_1985} Long-range
dipolar interactions in a back-gated two-dimensional electron gas
(2DEG) have been
predicted~\cite{spivak_phases_2004,spivak_colloquium_2010} to lead
to the existence of ``microemulsion'' phases intervening
the Fermi liquid and the Wigner crystal phase. Similar
microemulsion phases may appear in 2DEGs subject to magnetic
fields such that several Landau levels are
occupied.~\cite{koulakov_charge_1996,moessner_exact_1996,fogler_stripe_2002,eisenstein_insulating_2002}
\begin{figure}[ht]
\begin{center}
 \includegraphics[width=0.45\textwidth]{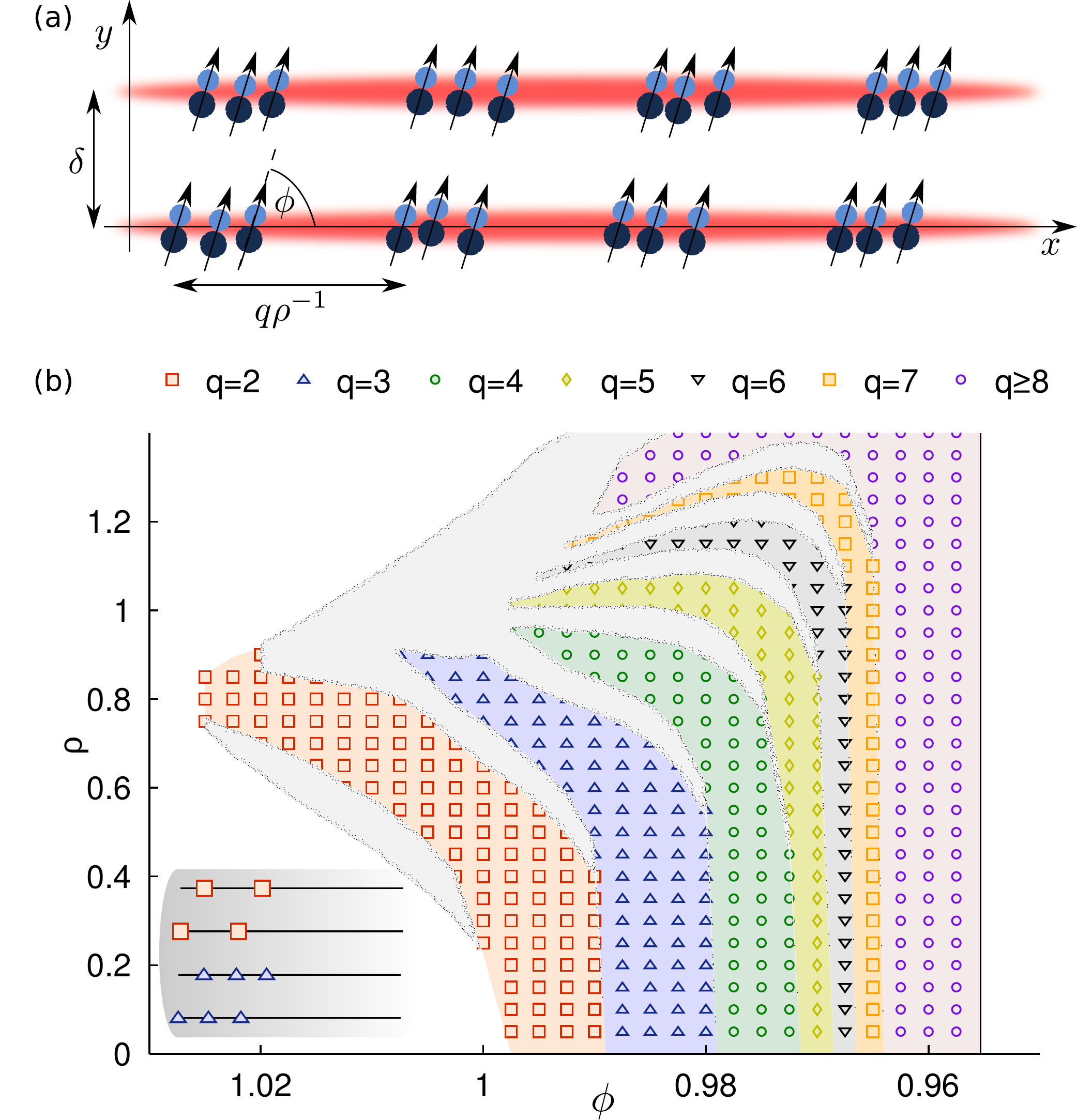}
\end{center}
\caption{\label{fig:pd} (Color online) Proposed setup to observe
cluster formation of polar molecules \fc{a}. The classical phase
diagram for two tubes as a function of the tilting angle $\phi$
and the particle density $\den$ is shown in \fc{b}. Lobe shaped
phases consisting of clusters with $q$ particles per tube emerge.
The phase separated regions are indicated by the shaded layer
surrounded with speckles. Inset: Optimized cluster configuration
in one unit cell for $\phi=1.01$, upper graph, and $\phi=0.99$, lower graph,
at $\den=0.7$, corresponding to $q=2$ and $q=3$, respectively.
}
\end{figure}

Theoretically, quantum emulsion phases are challenging to analyze
since they involve structures at length scales ranging from the
inter-particle distance to mesoscopic scales. 
In contrast, the conventional tools 
of many-body physics 
are mostly geared toward two particle
correlations, such as paired states, magnetism and charge density
wave. Experimentally, quantum emulsion phases are not easy to
probe since transport measurements can only provide indirect
evidence about their existence. Realizing long-range interactions
with systems of cold polar
molecules~\cite{santos_bose-einstein_2000,baranov_superfluid_2002,
doyle_editorial_2004,baranov_theoretical_2008,carr_cold_2009,lahaye_physics_2009}
can allow to explore emergent emulsion phases in a highly
controllable setting. Moreover, in such systems few-body
bound states,~\cite{wunsch_few-body_2011} trimer liquid
phases,~\cite{dalmonte_trimer_2011} and bound solitons~\cite{bauer_dipolar_2012} have been predicted.
\begin{figure*}
\begin{center}
 \includegraphics[width=0.99\textwidth]{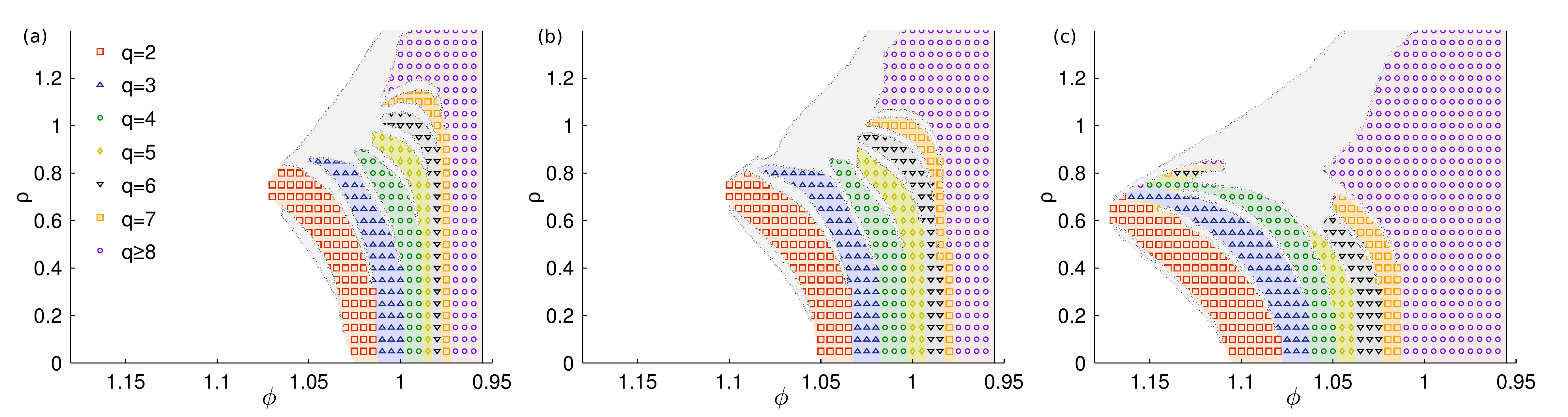}
\end{center}
\caption{\label{fig:pdL} (Color online) Phase diagram in the tilting angle
$\phi$ and the particle density $\den$ plane for \fc{a} $n_T=3$,
\fc{b} $n_T=4$, and \fc{c} $n_T=8$ tubes. The phase separation
regions are indicated by shaded layers. }
\end{figure*}

Here we demonstrate that the anisotropic and long-range character
of dipolar interactions leads to new types of clustered crystal
phases which appear at \emph{intermediate} values of the interaction
strength. Quantum mechanically, these phases melt into
distinct ``clustered Luttinger liquids'' characterized by the
decay of their density-density correlation functions. We calculate
the phase diagram and study the quantum melting of the clustered
phases when tuning the orientation of the dipoles. Our
calculations indicate that the clustered phases can be explored
under current experimental conditions. 

\section{Experimental setup}

We consider a setup in which polar 
molecules are confined to $n_T$ one-dimensional parallel tubes 
[see \figc{fig:pd}{a}], which can be realized by deep optical 
lattices.~\cite{chotia_long-lived_2012} The dipolar
moments are aligned in the plane of the tubes at an angle $\phi$
with respect to the tube direction. The inter-tube distance
$\delta$ is used as unit of length throughout this work.

The interaction energy between two molecules with dipolar moment
$\ve m$ is 
\begin{equation}
 V(r\ve e_r) = \frac{\mu^2 - 3 (\ve m \ve e_r)^2}{r^3}\;,
 \label{eq:dip}
\end{equation}
where $r\ve e_r$ is the inter-molecule displacement and
$\ve m= \mu (\cos \phi, \, \sin \phi, \, 0)^T$. For tilting angles
below the critical angle $\phi_c=\arccos 1/\sqrt{3}$, the
interaction between molecules in the same tube is attractive and
the system is unstable. Thus, we focus on dipolar orientations
$\pi/2 \geq \phi \geq \phi_c$ where the intra-tube interaction is
repulsive. Yet, molecules in different tubes attract when their
displacement along the tubes is not too large. It is precisely
this interplay between attraction and repulsion which leads to
the formation of clusters.

\section{Classical limit}

We first discuss
the emergence of mesoscopic structures in the classical limit ($\hbar\rightarrow 0$).

When the dipoles are oriented perpendicular to the tubes
($\phi=\pi/2$) the ground state is a Wigner crystal with $n_T$ molecules
per unit cell and periodicity $\den^{-1}$, where $\den$ is the linear
density of molecules.
Upon tilting the direction of the dipoles toward the critical
angle $\phi_c$, phases with an increasingly complex unit cell are formed, before
eventually becoming unstable to collapse at $\phi=\phi_c$.
In these phases, the unit cell consist of $q$ particles per tube
forming a superlattice with periodicity $q \den^{-1}$. For example, a phase with $n_T=2$
and $q=3$ is illustrated schematically in \figc{fig:pd}{a}.

We derived the phase diagram as a function of the tilting
angle $\phi$ and the density $\den$, by minimizing the classical
ground state energy with respect to the position of the molecules,
allowing for arbitrary periodic structures with up to $q=8$
molecules per unit cell in each tube. The phase diagram for
$n_T=2$ is shown in \figc{fig:pd}{b}. At small densities we
observe transitions to phases with monotonically increasing $q$ when
decreasing the tilt angle from $\pi/2$ toward $\phi_c$. Phases of
a fixed value of $q$ have a lobe-like structure, which bend with
increasing density toward larger $\phi$.
Quite generally phases in \figc{fig:pd}{b} terminate by
phase separated regions, indicated by a shaded layer surrounded with
speckles. The phase separated regions are determined by the Maxwell construction
which is applicable when the interfacial energy is positive.~\cite{spivak_phases_2004}
It is possible that phases with $q>8$, not captured by our present
calculation, are favorable in some
parts of the phase diagram. In particular, this is the case
very close to $\phi=\phi_c$ where we find that $q=8$ has the
lowest energy.

The origin of the cluster formation can be easily understood
by considering the case $\phi=\phi_c$. Then, the intra-tube
repulsion is precisely zero. In order to maximize inter-tube
attraction, it is favorable to form a single cluster with a
macroscopic number of particles,
corresponding to a $q\rightarrow \infty$ phase. As the angle is
tuned toward $\phi_c$, there must be either an infinite sequence
of transitions to increasingly higher values of $q$, or a
macroscopically phase separated region.~\cite{supp}

Next, we discuss systems with more than two tubes. Results for
$n_T=3$, $4$, and $8$ tubes are shown in \fig{fig:pdL}. The phase
diagrams for $n_T>2$ have a similar lobe structure as in the
$n_T=2$ case. The main difference is that with increasing number
of tubes the lobes extend to higher values of the tilting angle
$\phi$. Thus, clustered phases might be easier to observe in
systems with a larger number of tubes. As in the $n_T=2$ case,
phase separated regions appear between phases of
different $q$.

A two-dimensional system which consists of an infinite number of
tubes with dipoles aligned in the plane exhibits similar physics: the $q=1$ Wigner crystal
phase becomes locally unstable for $\phi>\phi_c$.
However, in this case, trial configurations with an increasing $q$
have monotonically lower energy (we have tried structures with up
to $q=128$), indicating that the ground state may be phase separated. 
In the low density limit, the dipoles form infinitely long
strings, which are mutually attractive~\cite{supp} and thus one
can show that the system is unstable to macroscopic phase
separation. Note that, for in-plane dipoles, the (logarithmically
divergent) surface energy is \emph{positive},~\cite{supp}
therefore macroscopic phase separation is possible (unlike the
out-of-plane case~\onlinecite{spivak_phases_2004}).

\section{Quantum mechanical analysis}

In quantum mechanical
systems with continuous translational symmetry, true long-range
crystalline order appears only in two dimensions or higher, even
at zero temperature. 
In one-dimensional systems, the density-density correlations decay
for large distances as a power law. Nevertheless, one can expect
that upon melting the clustered Wigner crystal phases by quantum
fluctuations, these phases will remain distinguishable 
by the nature of their quasi-long range correlations. We
term the resulting phases ``clustered Luttinger liquids.''

In a clustered Luttinger liquid phase, the slowest-decaying
component of the density-density correlations has a spatial period
of $\lambda = q \den^{-1}$. In a bosonized description, the
fundamental harmonic of the density operator is therefore of the
form $\cos [{2 \pi} (x + x_0) \den/q]$, where $x_0$ is a uniform
shift of the crystalline configuration. In terms of the ``counting
field'' $\phi(x)$~\cite{giamarchi_quantum_2004} defined relative
to the crystalline configuration we obtain for the bosonized
density
\[
 \rho(x) = \den - \frac{1}{\pi} \nabla \phi(x) + \den \cos\left[ \frac{2 \pi \den x }{q}+\frac{2 \phi(x)}{q}
 \right]+\dots,
\]
where the dots represent higher harmonics. The factor $q^{-1}$
present in the cosine alters the power law with which the density
correlations decays:
\begin{align}
 \langle \rho(x) \rho(0) \rangle &= \den^2 - \frac{K}{2\pi^2} \frac{1}{x^2} + \frac{\den^2}{2}\cos{\frac{2\pi \den x}{q}} \left( \frac{\alpha}{x} \right)^{\frac{2K}{q^2}}\;.
 \label{eq:nnLut}
\end{align}
Here, $K$ is the Luttinger parameter. Microscopic considerations
\cite{supp} suggest that $K\propto q$. Therefore, 
the exponent with which the 
density-density correlation function decays is proportional to
$1/q$.
Phases with larger $q$ thus have a slower decay of the
density-density correlation function, and are increasingly
``classical'' in
nature. 
\begin{figure}
\begin{center}
 \includegraphics[width=0.49\textwidth]{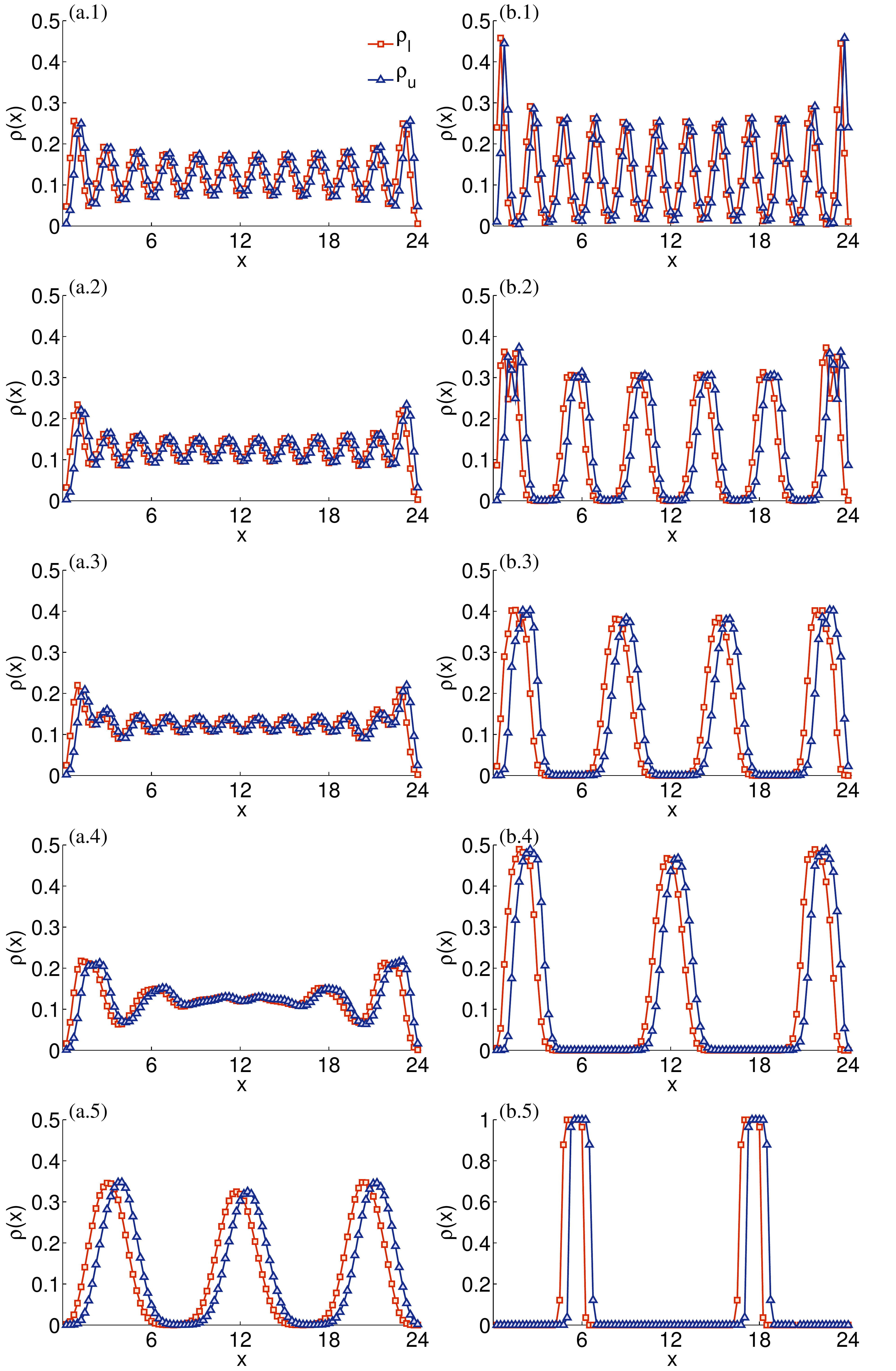}
\end{center}
\caption{\label{fig:nDMRG} (Color online) Particle density $\den(x)$ in the lower (red squares) and
the upper (blue triangles) tube for dipolar strength $\gamma=8$, left column, and $\gamma=50$, right column, 
for a system of length $L=24 \delta$, density $\den=0.5\delta^{-1}$ and lattice
spacing $a=\delta/4$. 
From top to bottom the tilt angle takes the values $\phi=\lbrace 1.02,\,0.99,\,0.98,\,0.97,\,0.96\rbrace$ where for $\gamma=50$ 
pronounced clusters with $q=\lbrace 1,\,2,\,3,\,4,\,6\rbrace$ are found, respectively. 
The data is obtained by DMRG for systems with open boundary conditions.}
\end{figure}%

In order to make quantitative predictions about the phase diagram
in the presence of quantum fluctuations, we have investigated
a system of two tubes numerically
by means of density matrix renormalization group
(DMRG)~\cite{white_density_1992,schollwoeck_density-matrix_2005}
simulations.
To this end, we introduce the lattice Hamiltonian
\begin{align}
   \hat H &= -t \sum_{\alpha,i} \left[c_{\alpha, i}^\dag c_{\alpha, i+1}^\nag +c_{\alpha, i+1}^\dag c_{\alpha, i}^\nag\right]
  \nonumber \\
  &+ \frac{\mu^2}{\delta^3} \sideset{}{'}{\sum}_{i,j,\alpha,\beta} V_d[(i-j)a/\delta,\,\alpha-\beta] \hat{n}_{\alpha, i} \hat{n}_{\beta, j}\;,
 \label{eq:ham}
\end{align}
where $c_{\alpha, i}^\nag$ ($c_{\alpha, i}^\dag$) destroys
(creates) a particle at site $i$ of tube $\alpha=1,2$, and
$\hat{n}_{\alpha,i}=c_{\alpha, i}^\dag c_{\alpha, i}^\nag$ counts the
number of particles. Due to the strong on-site repulsion we treat the particles 
as hard-core, and therefore for the quantities we compute here (\eg, density
distributions and ground-state energies), it does not matter whether the
particles are bosons or fermions.
The discrete Hamiltonian can represent a continuous system by
taking the lattice spacing $a\rightarrow 0$ while keeping the
product $ta^2 = {1}/({2m})$ constant, where $t$ is the hopping
strength and $m$ is the mass of the particles in the continuum.
The primed sum indicates that the singular contribution where
$i=j$ and simultaneously $\alpha=\beta$ is omitted. The dipolar
energy $V_d(x,y)$ is given by \eq{eq:dip} with $\ve r=(x,\,y)^T$.

It is convenient to introduce the dimensionless quantity $\gamma$,
which is the ratio between the typical dipolar interaction energy
$E_{\text dip}$ and the typical kinetic energy $E_{\text kin}$.
These energies can be estimated as $E_{\text dip} \sim \mu^2
\den^3$ and $E_{\text kin} \sim {\den^2}/{m}$, respectively, and
thus $\gamma \sim \mu^2 \den m$. In the limit of strong
interactions, $\gamma\gg 1$, the system is expected to be
essentially classical, and the phase diagram is expected to be
similar to that of \fig{fig:pd} with the Wigner crystalline phases
replaced by clustered Luttinger liquids. Conversely, for
$\gamma\ll 1$, quantum fluctuations dominate, and we expect only
the $q=1$ phase to survive.

The particle density $\den(x)$ evaluated with DMRG for the
Hamiltonian \eqw{eq:ham} exhibits clear signatures
of clustered phases, see \fig{fig:nDMRG}. We consider
a system of finite length $L=24\delta$ with open boundary
conditions and particle density $\den=0.5\delta^{-1}$ and
dipolar strength $\gamma=50$. The lattice
constant is $a=\delta/4$; no significant change in the results was
observed when $a$ was decreased to $\delta/6$. Since the reflection symmetry about a
plane perpendicular to the tube is broken for any tilting angle
except $\phi=\pi/2$, the density of the upper $\den_u$ and the
lower $\den_l$ tube are slightly shifted. Additionally, we observe
that the height of the peaks in the density decreases toward the trap
center, consistent with quasi-long range order.
Remarkably, the rate of the decay decreases strongly with
increasing $q$, as expected from the Luttinger liquid
analysis, \eq{eq:nnLut}.
\begin{figure}
\begin{center}
\includegraphics[width=0.45\textwidth]{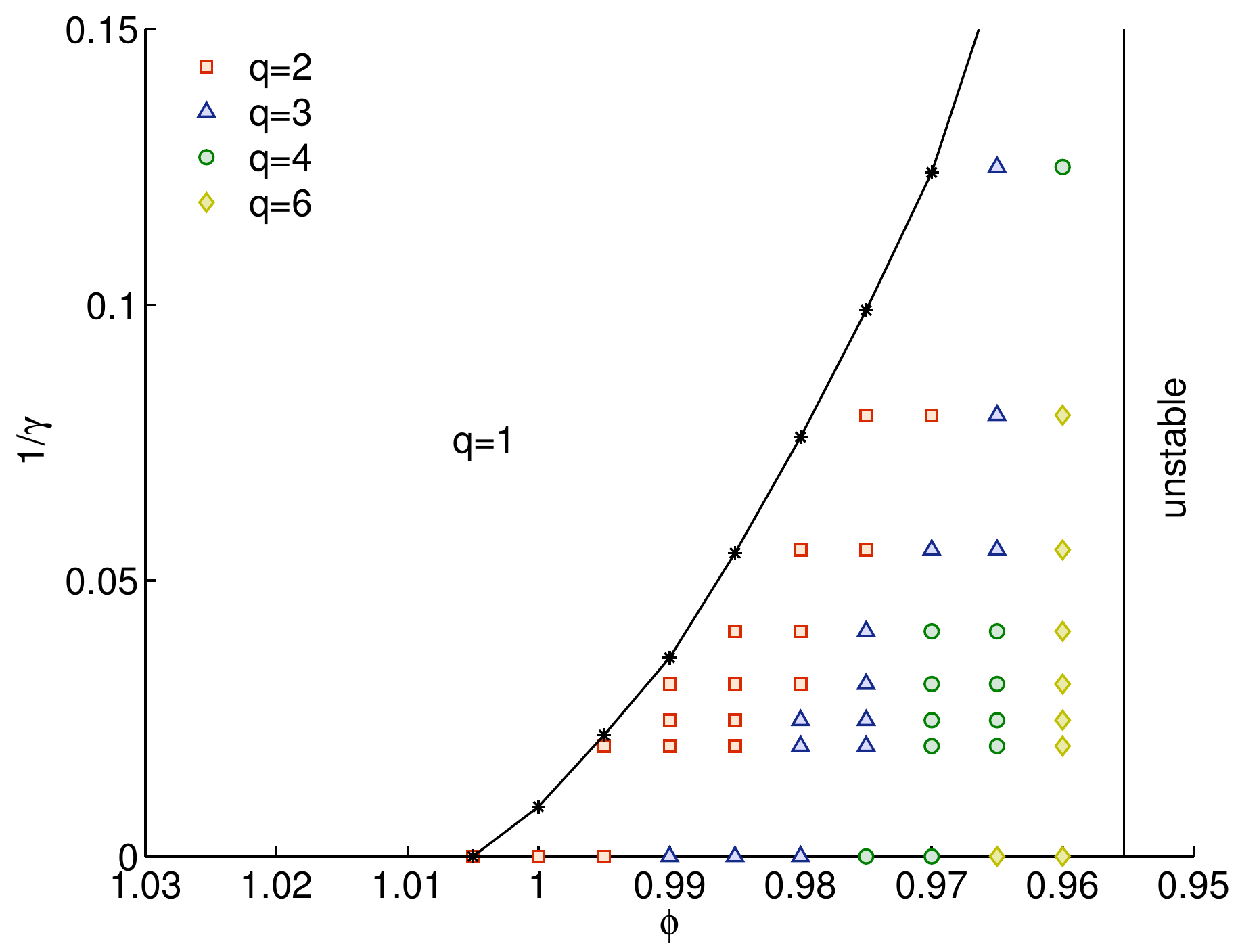}
\end{center}
\caption{\label{fig:pdDMRG} (Color online) Quantum phase diagram for a two tube
system of length $L=24\delta$, density $\den=0.5\delta^{-1}$, and
lattice spacing $a=\delta/4$, as a function of the tilting angle
$\phi$ and the ratio between the kinetic and the interaction
energy $1/\gamma$. Right of the asterisks connected by lines
cluster formation can be observed. To the left of this line, the
ground state is a $q=1$ Luttinger liquid.}
\end{figure}

The complete quantum phase diagram for the two tube system
as a function of the tilting angle $\phi$ and
the ratio between kinetic and interaction energy
$1/\gamma$ is shown in \fig{fig:pdDMRG}.
For $1/\gamma=0$, the results were obtained by classical
minimization of the interaction energy. The DMRG simulations are
used to extend the results to $1/\gamma>0$.
The phases are determined from the density distribution by
calculating the number of particles localized within one cluster.
The clustered Luttinger liquid phases with $q>1$
extend to considerably large values of $1/\gamma$, making the
realization of clustered phases feasible in experiments with
cold dipolar molecules.

\section{Experimental implications}

For typical densities of 
$\den=10^4 \text{cm}^{-1}$, $\gamma \sim \lbrace0.7, \,3.3, \,6.8, \,49.4, \,63.3\rbrace$ 
can be achieved in experiments with KRb, RbCs, NaK, NaCs, and LiCs, 
respectively. 
(See \onlinecite{supp} for details on this calculation.)

The density regime which is most favorable for observing clustered phases
is $\den\sim 0.5\delta^{-1}-\delta^{-1}$. For $\den\sim 10^4 \text{cm}^{-1}$, this
corresponds to an inter-tube separation of $\delta \sim 5\cdot10^{-5}
\text{cm}$, easily attainable using an optical potential created by a laser with wavelength
$\sim 1 \mu\text{m}$.

NaCs and LiCs are thus the most promising candidates to realize
clustered phases, due to their large dipolar moments. In order to
make the clustered phases more robust, one can add a shallow
periodic potential along the tubes. Such a periodic potential
quenches the kinetic energy, thus increasing the effective value
of $\gamma$. As a consequence cluster formation arises at much
weaker dipolar moment also attainable by KRb.

Other effects that can be important for experiments are (i)
the incommensurability of the particle number with the cluster
size, (ii) the shallow trap potential along the tube direction,
(iii) the strong but finite transverse confinement, (iv) quantum
fluctuations in the orientation of the dipoles, and (v) finite
temperature effects. Cluster formation is extremely stable with
respect to (i) and (ii).~\cite{supp} Incommensurability leads to a
slight rearrangement of clusters and the consequence of the
shallow trap along the tubes is merely that the distance between
the clusters is reduced. One-dimensional tubes are realized by 
strong transverse confinement potentials (iii). Therefore, 
we consider the interactions computed for molecules with 
transverse wavefunctions, corresponding to a parabolic confinement, 
and compare them to the bare, one-dimensional interactions. The 
renormalization of the interactions due to the transverse confinement
can be evaluated from a multipole expansion yielding $\Delta E^\perp_{\text{dip}}/E_{\text{dip}}
\lesssim \sigma^2/\delta^2$, where $\sigma$ is the 
spread of the wave function in transverse direction.~\cite{supp} Under standard
experimental conditions $\sigma\sim 25 \text{nm}$ and thus $\Delta
E^\perp_{\text{dip}}/E_{\text{dip}} \lesssim 0.0025$. This ratio
has to be compared with the relative energy difference between the
clustered $q>1$ and the uniform $q=1$ phase, which typically is
$0.2-0.5 \gg \Delta E^\perp_{\text{dip}}/E_{\text{dip}}$.
Therefore, the renormalization of the interaction energy due to
the finite strength of the transverse confinement is no obstacle
for the observability of the clustered phases. The quantum
fluctuations of the dipoles around the orientation of the electric
field (iv) renormalize the dipolar potential by $\Delta
E^e_{\text{dip}}/E_{\text{dip}} \lesssim R_e/\delta \sim 0.001$, where
$R_e$ is the bond length of the molecule. Thus, this effect is also
small. The temperature scale (v) below which we expect
strong tendency toward cluster formation is proportional to the
dipolar energy $E_{\text dip}$. In units of the Fermi temperature
$T_F\sim E_{\text{kin}}$ the crossover temperature is
$T_{\text{cross}}= \alpha \gamma$, where the proportionality
constant $\alpha$ can be estimated from the relative energy
difference of the uniform and the clustered phases, \ie, $\alpha
\sim 0.2 - 0.5$. Depending on the molecule~\cite{supp}
the crossover temperature is $T_{\text{cross}} \gtrsim T_F$.

\section{Conclusions and Outlook}

In summary, the formation of clustered
phases of polar molecules is a distinct consequence of the
long-range and anisotropic nature of their interactions. We find
that the clustered phases can be explored under current
experimental conditions. A variety of techniques can be employed
to observe the clustered phases, including
elastic light scattering,~\cite{mueller_weak_2001} noise correlations in
time-of-flight images,~\cite{altman_probing_2004} and optical
quantum nondemolition detection.~\cite{wunsch_few-body_2011,zinner_few-body_2011}
Cluster formation should also contribute additional dissociation
resonances in lattice modulation
experiments~\cite{stoeferle_transition_2004} and rf spectroscopy.~\cite{regal_measurement_2003,gupta_radio-frequency_2003,regal_creation_2003}

A particular exciting direction for future research is to
study excitations in this system. For the simple $q=1$
phase, the unit cell consists of a single particle
per tube and thus only the acoustic mode exists. However, in clustered
crystal phases with $q>1$ also optical branches should exist
in the excitation spectra.

\begin{acknowledgments}
We are grateful to E. G. Dalla Torre for
insightful discussions about the Luttinger liquid analysis.
M.K. wants to thank W.~von der Linden and E.~Arrigoni for fruitful discussions.
The authors acknowledge support from Harvard-MIT CUA, the
NSF Grants No. DMR-07-05472 and No. DMR-07-57145, the DARPA OLE program,
AFOSR Quantum Simulation MURI, AFOSR MURI on Ultracold Molecules,
the Austrian Science Fund (FWF) under Project No. P18551-N16 (M.K.) and
within the SFB ViCoM (F41) (M.G.), as well as the Austrian Marshall
Plan Foundation (M.K.). Calculations have been partly performed on
the iCluster of Graz University of Technology.
\end{acknowledgments}

\appendix

\section*{Supplementary Material}

\textit{Incommensurability and longitudinal confinement.---}Here, we investigate
in detail the effects of the particle number incommensurability with the
cluster size and of the shallow confining potential along the tube direction.

\begin{figure}
\begin{center}
 \includegraphics[width=0.4\textwidth]{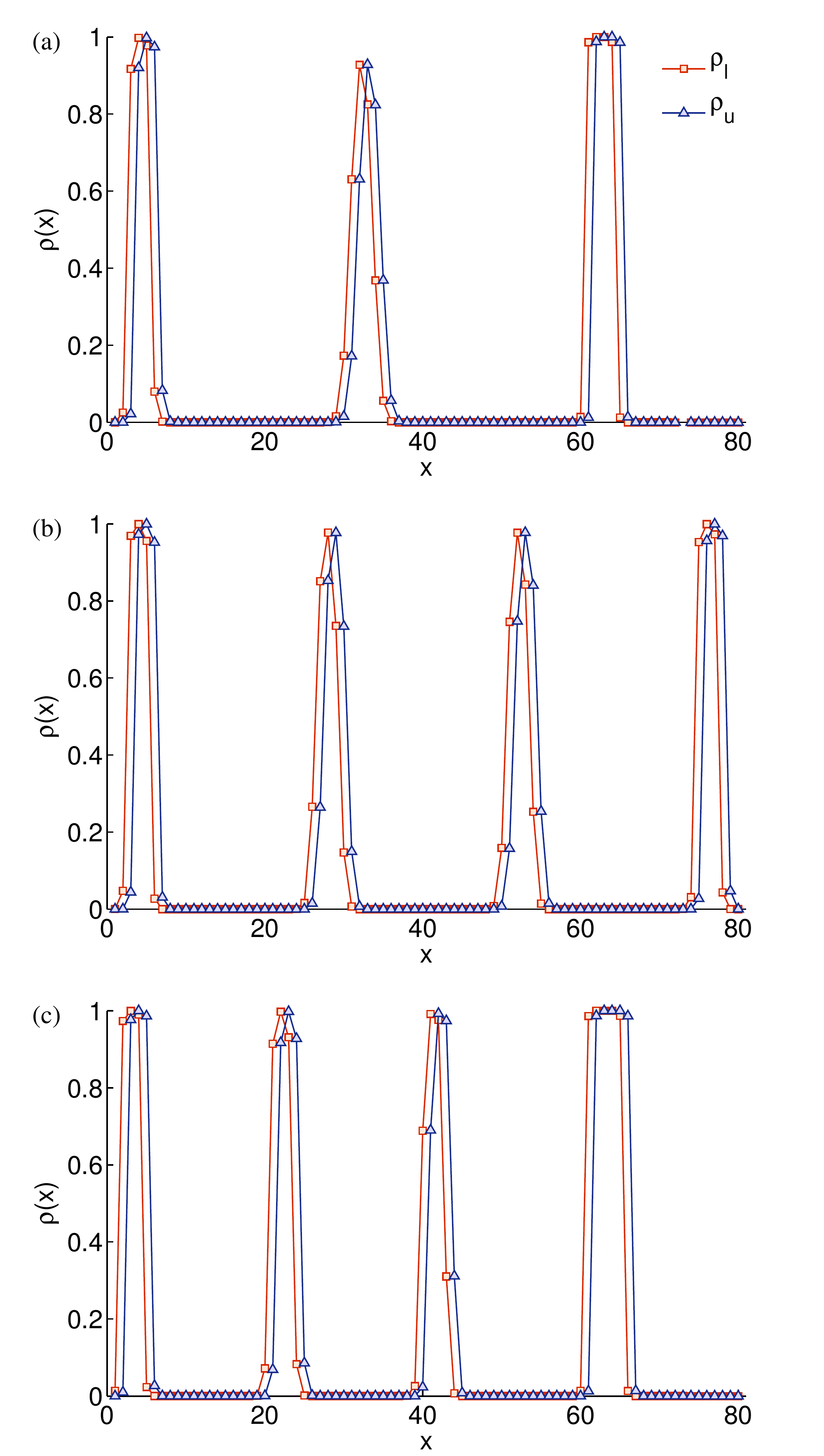}
\end{center}
\caption{\label{fig:nDMRGIn} (Color online) Particle density $\den(x)$ in the lower (red squares)
and in the upper (blue triangles) tube for the ratio of the interaction to
kinetic energy $\gamma=2.4$, tilt angle $\phi=1$,
length $L=80\delta$, and lattice constant $a = \delta$ for systems with \fc{a} $N=10$,
\fc{b} $N=12$, and \fc{c} $N=14$ particles per tube. The cluster formation is insensitive
as to particle densities which are incommensurable with the cluster size.}%
\end{figure}

To demonstrate that the cluster formation is
stable against the incommensurability of the particle number with the cluster size, we
evaluate the density distribution for systems with $N=10$, $N=12$, and $N=14$ polar
molecules per tube, respectively. The remaining system parameters are: ratio of interaction and
the kinetic energy $\gamma=2.4$, tilt angle $\phi=1$, length $L=80\delta$, and lattice
constant $a=\delta$.
The cluster formation is commensurable with $N=12$ particles, where four clusters with $q=3$ particles
are found, see \figc{fig:nDMRGIn}{b}.
However, both $N=10$ and $N=14$ are incommensurable with the $q=3$ phase.  Regardless, we
observe a pronounced cluster formation, see \figcc{fig:nDMRGIn}{a}{c}.
\begin{figure*}
\begin{center}
 \includegraphics[width=0.95\textwidth]{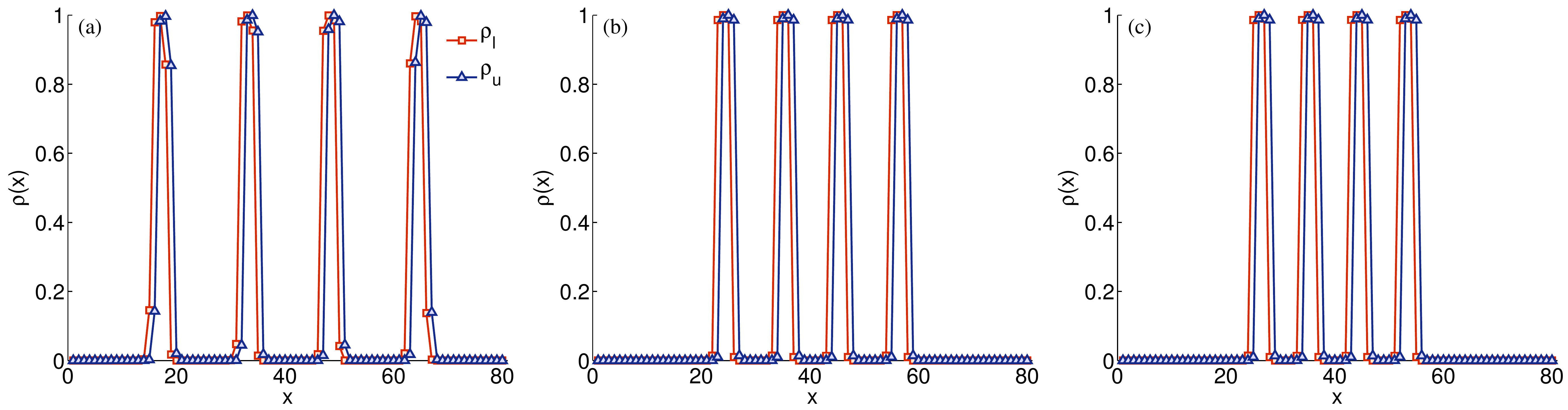}
\end{center}
\caption{\label{fig:nDMRGTr} (Color online) Particle density $\den(x)$ for the same system as described
in the caption of \fig{fig:nDMRGIn} but with $N=12$ throughout and a confining potential of strength
\fc{a} $\mathcal{C}=0.1t$, \fc{b} $\mathcal{C}=1.0t$, and \fc{c} $\mathcal{C}=4.0t$. The cluster formation is
robust with respect to the trap potential. The mere effect of the trap in the studied regime
of the confining strength is that the distance between the clusters is reduced.}%
\end{figure*}%

Next, we study the effects of the harmonic confining potential along
the tube direction,
which we incorporated into the Hamiltonian by a variable on-site
energy $\epsilon_j = \mathcal{C} (ja-L/2)^2/L^2$. It turns out that the
formation of clusters is also very insensitive to the confining
potential. Results for the particle distribution are shown in
\fig{fig:nDMRGTr} for \fc{a} $\mathcal{C}=0.1t$, \fc{b} $\mathcal{C}=1.0t$, and \fc{c}
$\mathcal{C}=4.0t$. For the considered values of $\mathcal{C}$, the main effect of the
harmonic confinement potential is that the distance between
clusters of particles is reduced as compared to the uniform case
shown in \figc{fig:nDMRGIn}{b}.

\textit{Multipole expansion.---}The renormalization of the 
interactions due to the fluctuations of the dipoles around
the orientation of the applied electric field and due to 
the finite spread of the transverse wavefunctions, can be 
evaluated from a multipole expansion of the electric potential
\begin{equation}
 \phi(r) = \sum_{lm}  \frac{q_{lm}}{2l+1} \frac{Y_{lm}}{r^{l+1}}\;,
 \label{eq:multi}
\end{equation}
where $q_{lm} = \int Y_{lm}^* r^l \rho(\ve r) d^3 \ve r $ 
with the charge distribution $\rho(\ve r) = \rho_+(\ve r-\ve r_e) -
\rho_-(\ve r+\ve r_e)$. 

The largest effect of this renormalization concerns dipoles
which are separated roughly by $\delta$. The electric field induced 
from one dipole at the location of the second dipole is
\[
 \ve E(\delta \ve e_x) = \nabla \phi(\delta \ve e_x) \;,
\]
and the potential energy is given by $V=\ve m \ve E (\delta \ve e_x)$. 
The leading order term is clearly the dipole-dipole interaction 
$E_{22} = \ve m \ve E_{2} (\delta \ve e_x)$. 

The next order is the dipole-quadrupole interaction $E_{42}$, induced from 
the fluctuations of the dipoles around the electric field. This 
contribution is nonvanishing, since heteronuclear polar molecules are
considered. From the multipole expansion \eqw{eq:multi} we infer the 
microscopic quadrupolar renormalization
\[
 \frac{E_{42}}{E_{22}} = \text{const.}  \frac{R_e}{\delta} \;,
\]
with $\text{const.}\lesssim 1$ and $R_e$ the bond length of the molecule.

As the transverse confinement obeys rotational symmetry, there is no
quadrupole contribution from the transverse wavefunction. Therefore,
the first non-vanishing contribution due to the spread of the transverse 
wavefunction is of the dipole-seipole (6 charges) form $E_{62} = \ve m \ve E_6 (\delta \ve e_x)$
\begin{equation}
 \frac{E_{62}}{E_{22}} \lesssim \frac{\sigma^2}{\delta^2} \;,
 \label{eq:deisei}
\end{equation}
where we used $\rho_+(\ve r)=\rho_-(\ve r) = q/(\pi \sigma^2) 
\exp[-(y^2+z^2)/\sigma^2] \delta(x)$ and thus $q_{lm} \sim q R_e \sigma^{l-1}$. 

\begin{figure}
\begin{center}
 \includegraphics[width=0.48\textwidth]{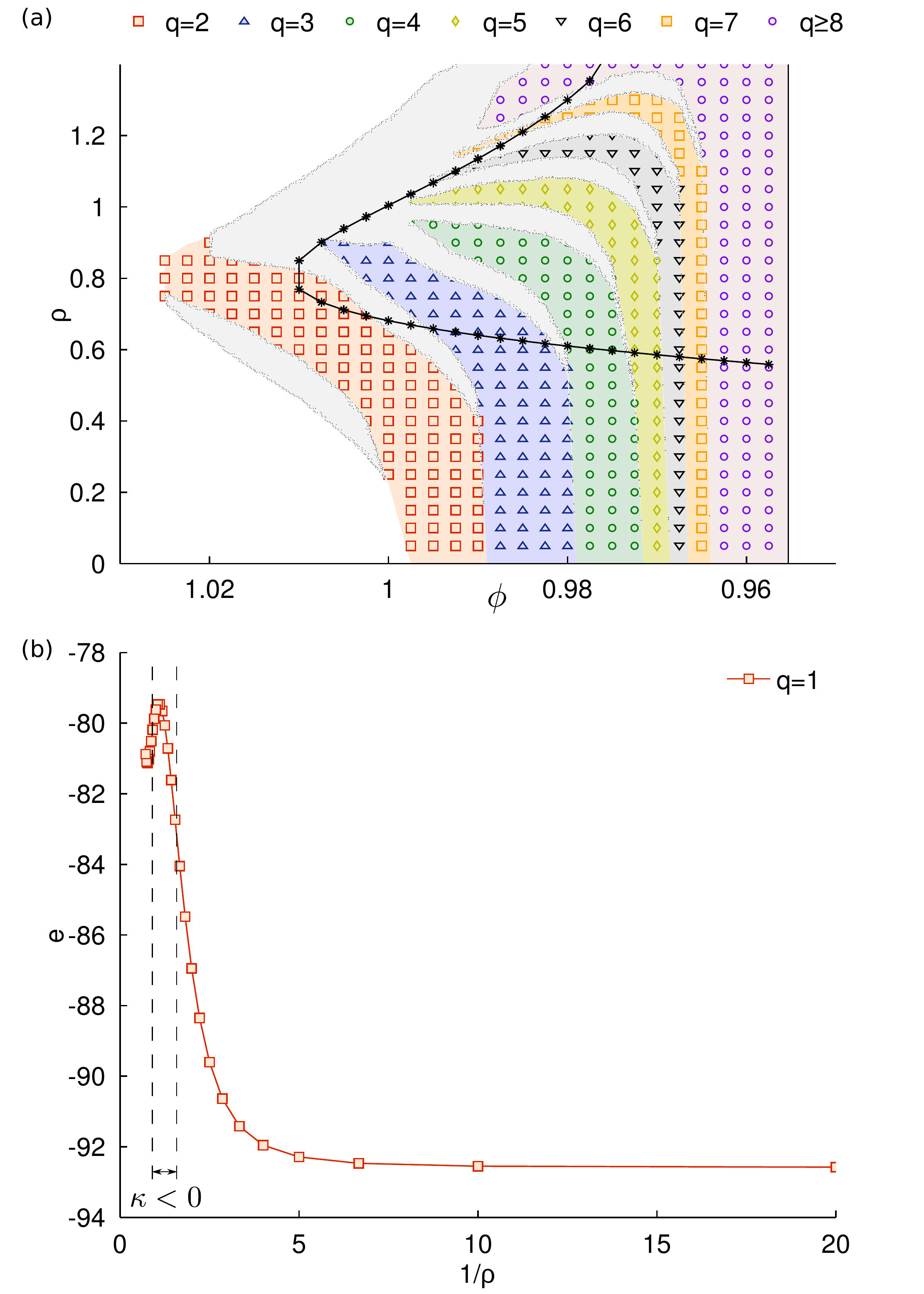}
\end{center}
\caption{\label{fig:spinodal1D} (Color online) \fc{a} Phase diagram of dipolar molecules confined in
two tubes as in \fig{fig:pd} and spinodal line for the $q=1$ phase, asterisks connected by lines, which separates
regions of positive and negative compressibility. \fc{b} Ground state energy of the
$q=1$ phase for $\phi=0.99$. In the region between the dashed vertical lines the curvature
of the energy per particle is negative indicating the density regime where the uniform phase is
certainly unstable.
}
\end{figure}

\textit{Local stability of clustered phases.---}One concern could
be that the uniform $q=1$ crystalline phase may be metastable, and
therefore the system might not be able to find its ground state on
experimental timescales. Thus, it is useful to explicitly evaluate
the stability of the uniform $q=1$ phase. A system is locally
unstable if its compressibility $\kappa$ is negative. The
compressibility can be evaluated from $\kappa^{-1} = \partial^2
\varepsilon /
\partial (\den^{-1})^2$, where $\varepsilon$ is the energy per particle
and $\den$ the particle density. Thus
it is useful to evaluate the spinodal line, which separates the
regions with positive and negative compressibility, see asterisks
connected by lines in \figc{fig:spinodal1D}{a}. On the right hand
side of the spinodal line, close to the critical angle $\phi_c$,
there is a large region
where the uniform
$q=1$ phase is locally unstable.
In fact, it can be shown that any clustered phase (with an
arbitrary $q$) must become locally unstable for a sufficiently
large density and sufficiently close to $\phi=\phi_c$.

Assume that $\phi=\phi_c$, such that the intra-tube interaction is
zero.
In the high density limit, the energy per particle is proportional
to the particle density times the integrated inter-tube energy.
Explicitly integrating the inter-tube interaction gives
$e=-{\mu^2 \den}\sin^2(\phi)/{\delta^2}$.
Therefore $\kappa^{-1} < 0$.
As we move away from the critical angle,
the repulsive intra-tube interaction gives a contribution proportional to $(\phi-\phi_c)\den^3$.
Adding these two contributions and calculating the
compressibility, we get that the critical density
$\den^\star(\phi^\star)$ at which the phase becomes unstable
satisfies $\den^\star \propto (\phi^\star-\phi_c)^{\frac{1}{2}}$.

This argument captures the main characteristics of the spinodal
line at high densities. Below a certain density, the spinodal line
terminates and the phase becomes locally stable, as can be
seen in \figc{fig:spinodal1D}{a} for $q=1$. In \figc{fig:spinodal1D}{b}
we show the calculated ground state energy of the $q=1$ phase for
$\phi=0.99$, showing the region in
which $\kappa<0$.
The relative extend and shape of the spinodal line for systems
with a larger but finite number of tubes, as shown in \fig{fig:pdL}, 
is similar to the case of two tubes.

At the critical angle $\phi=\phi_c$ the ground state consists
of dipoles which are arbitrarily close to each other corresponding
to a $q = \infty$ phase. From this and from the analysis of the
compressibility follows that, there must be either an infinite sequence of
phases with increasing $q$, or a region of macroscopic phase separation
between a phase of finite $q$ and vacuum upon approaching $\phi=\phi_c$.

\textit{Infinite number of tubes.---}Next, we study the phase
diagram in the case of an infinite number of tubes. We find that
the $q=1$ phase is locally unstable in a region $\pi/2>\phi>\phi_c$.
Our results are consistent with macroscopic phase
separation between the $q=1$ phase and vacuum. So unlike the case
with a finite number of tubes, intermediate phases with $q>1$ are
not realized.

The classical ground state energy of the two-dimensional system is
evaluated by Ewald summation techniques.\cite{ewald_berechnung_1921,essmann_smooth_1995,grzybowski_ewald_2000}
We allow for a unit cell containing $q$ particles. The unit cell
is a parallelogram of length $q\den^{-1}$, height $\delta$, and an
arbitrary angle. In addition, for $q>1$, we assume that the
particles within a unit cell are equally spaced. The angle and the
intra-unit cell spacing are treated as variational parameters. The
same scheme was used to optimize the ground state energy in the
case of a finite number of tubes, and gave excellent results
compared to a full optimization with respect to the positions of
all the particles in the unit cell presented in the main text [see also inset of
\figc{fig:pd}{b}].

\begin{figure}
\begin{center}
 \includegraphics[width=0.48\textwidth]{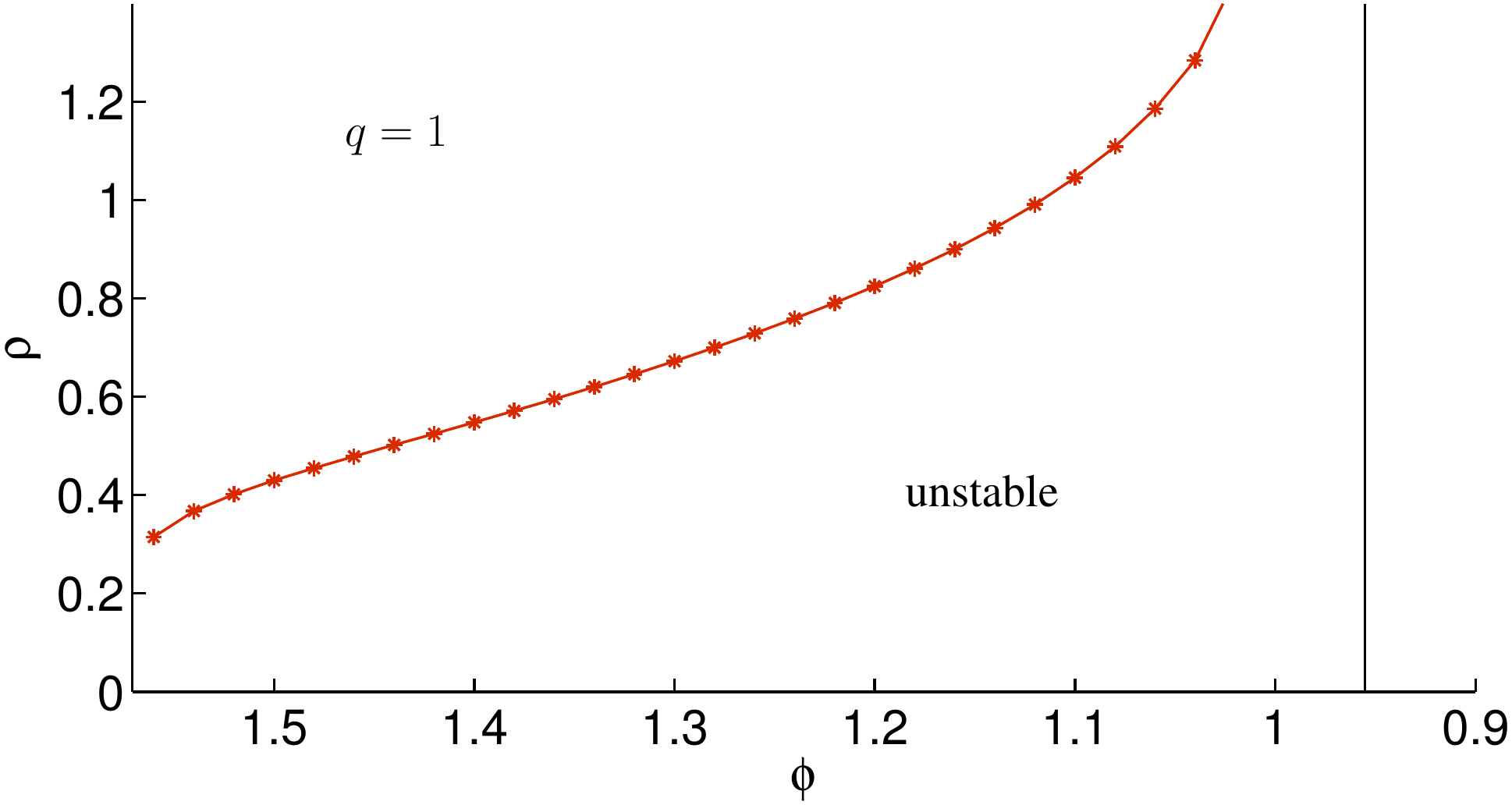}
\end{center}
\caption{\label{fig:spinodal2D} (Color online) Spinodal line for a system with infinitely many
tubes, which indicates the region of instability of the $q=1$ Wigner crystal phase.
}
\end{figure}

\fig{fig:spinodal2D} shows the spinodal line for the $q=1$ phase,
at which the compressibility is zero. To the right of this line,
the compressibility is negative, and the $q=1$ phase is locally
unstable. In the unstable regime, phases with $q>1$ are lower in
energy than the $q=1$ phase; however, the energy decreases
monotonically with increasing $q$, up to the largest $q$ that we
tried ($q=128$), and is always higher than a phase separated state
with macroscopic regions of $q=1$ and vacuum.
In the low density limit strings of polar molecules are
formed, whose mutual interaction is attractive. Therefore, at
sufficiently low densities the system always phase separates.

In two-dimensional systems with dipolar interactions, the surface
tension between two phases of different density diverges
logarithmically.~\cite{spivak_phases_2004} For dipoles pointing out
of the plane, the divergent term is negative, and hence
macroscopic phase separation is always unstable toward forming
``microemulsion'' phases with emergent, mesoscopic structures.
For in-plane dipoles, however, the divergent energy is
\emph{positive}, and therefore macroscopic phase separation is
possible.

To demonstrate this, we consider a system with a linear domain
wall between two phases $A$ and $B$. The dipolar contribution to
the total energy is
\begin{equation}
 E=\frac{1}{2} \int d \ve r \int d\ve r'  \den(\ve r) \den(\ve r') V(\ve r - \ve r')\;,
 \label{eq:eMac}
\end{equation}
where $\den(\ve r)$ is the density at position $\ve r$, which is $\den_A$ in phase $A$ and $\den_B$ in phase $B$.
Alternatively, the energy can be written as
\begin{equation}
 E=V_A e_A + V_B e_B + E_\sigma \;,
 \label{eq:eAdd}
\end{equation}
where $e_A (e_B)$ is the energy density of the homogeneous $A(B)$
phase, $V_A(V_B)$ is the area of phase $A (B)$, and
$E_\sigma$ the surface energy between the two phases. The bulk
contributions are defined as
\[
 V_{A(B)} e_{A(B)} = \frac12 \int_{\ve r \in \Omega_{A(B)}} d \ve r \int   d \ve{r}' \den_{A(B)}^2 V(\ve r - \ve r') \;,
\]
where $\Omega_{A(B)}$ defines the domain of phase $A(B)$. The integral over $\ve r'$ is taken
over the entire plane $\mathbbm{R}^2$. Comparing
\eqq{eq:eMac}{eq:eAdd} yields the surface energy
\[
 E_\sigma = -\frac12 \int_{\Omega_{A}} d\ve r\int_{\Omega_{B}} d\ve r' (\den_A-\den_B)^2 V(\ve r - \ve
 r')\;.\label{eq:Esigma}
\]
For dipolar interactions, $E_\sigma/L$ (where $L$ is the linear
dimension of the system) diverges as $\log(L)$ as
$L\rightarrow\infty$. In the following, we determine the sign of
$E_\sigma$ for a straight
domain wall
oriented along the $y$-direction.
In order to make contact to the results of
Ref.~\onlinecite{spivak_phases_2004} where the dipoles are
pointing out of plane, we consider a general dipole orientation:
\[
 \ve m = \mu ( \cos \phi \cos \theta,\; \sin \phi \cos \theta,\;\sin \theta)^T\;,
\]
where the angle $\theta$ is measured from the plane.
An explicit calculation reveals that the surface energy vanishes
at an angle $\theta_c$ given by
\[
 \theta_c = \arccos \sqrt{\frac{1}{1+\cos^2\phi}}\;.
\]
For $\theta>\theta_c$, $E_\sigma < 0$, and macroscopic phase
separation is impossible; for $\theta < \theta_c$, $E_\sigma > 0$
and phase separation is allowed. In particular, for
$\theta=\pi/2$, $E_\sigma<0$ (consistently with Ref.
\onlinecite{spivak_phases_2004}), whereas for $\theta=0$, $E_\sigma
\ge0$ for any value of $\phi$.

In summary, in the case of a two-dimensional array of tubes, it
seems that the $q=1$ phase terminates at a first-order transition
to vacuum, and no higher $q$ phases exist.

\textit{Numerical values for $\gamma$.---}The numerical data used 
to evaluate the ratio between the dipolar and the kinetic 
energy $\gamma$ is summarized in \tab{tab:dipolar}.
\begin{table}
  \caption{ Numerical values for the ratio between the dipolar and the kinetic energy $\gamma$ for dipoles $AB$ and the quantities which are necessary to estimate this ratio. The dipole strength is denoted as $\mu$ and $m_A$ ($m_B$) is the mass of constituent $A$ ($B$). We assume the linear density to be
$\den = 10^4 \text{cm}^{-1}$. }
  \label{tab:dipolar}
  \centering
  \begin{tabular}{crrrrr}\hline \hline
      $\quad$ AB$\quad$ & $\qquad \mu/De$ & $\qquad m_A/u$ & $\qquad m_B/u$ & $\qquad \qquad \gamma$ \\
       \hline KRb~\cite{ni_high_2008} &$0.6$  &$39.1$ &$85.5$ &$0.7$ \\
RbCs~\cite{debatin_molecular_2011} &$1.0$   &$85.5$ &$132.9$ & $3.3$ \\
NaK~\cite{park_quantum_2011} & $2.7$ & $23.0$ & $39.1$ & $6.8$  \\
NaCs~\cite{haimberger_processes_2006} &$4.6$ & $23.0$ & $132.9$ & $49.4$ \\
LiCs~\cite{deiglmayr_formation_2009} &$5.5$ &$6.9$  &$132.9$ &$63.3$ \\
  \hline \hline
  \end{tabular}
\end{table}

\textit{Luttinger parameter $K$.---}In order to reveal the
dependence of the Luttinger parameter $K$ of the clustered system
on $q$, we relate it to the Luttinger parameter $K_0$ of the
uniform ($q=1$) system. The low-energy properties of a $q=1$
liquid of polar molecules in one-dimensional tubes are described
by~\cite{giamarchi_quantum_2004,kollath_dipolar_2008}
\begin{align*}
 H_{q=1}&=\int dx \rho_0 \frac{(\nabla \Theta)^2}{2m} + \frac{\mu^2 \rho_0^2}{\pi^2} (\nabla \phi)^2 \\
 &=\frac{u_0}{2\pi} \int dx [ K_0 (\nabla \Theta)^2+ \frac{1}{K_0} (\nabla \phi)^2]\;,
\end{align*}
where we introduced the microscopic Luttinger parameters $K_0$ and
$u_0$ in the second line. The potential energy contribution is
given by the particle density $\rho$ times the dipolar energy per
particle $\sim \mu^2\rho^3$. This gives in total the contribution
$\sim \mu^2\rho^4 $ to the energy density. Expanding the density using $\rho = \rho_0 +
\delta \rho$ up to second order in $\delta \rho$ gives $\rho^4
\sim \rho_0^2 \delta \rho^2 = \rho_0^2 (\nabla \phi/\pi)^2$. The
clustered state of polar molecules has the same kinetic energy but
a reduced potential energy
\begin{align*}
 H_q&=\int dx \rho_0 \frac{(\nabla \Theta)^2}{2m} + \frac{\mu^2 \rho_0^2}{\pi^2 q^2} (\nabla \phi)^2 \\
 &=\frac{u}{2\pi} \int dx [u K (\nabla \Theta)^2+ \frac{u}{K} (\nabla \phi)^2]\;,
\end{align*}
as the potential energy is now given by the density of clusters
$\rho/q$ times the interaction energy of a single cluster
$(q\mu)^2/(q/\rho)^3$. This gives rise to the additional factor
$1/q^2$. Comparing the Luttinger parameters in the $q=1$ and $q>1$
cases gives
\[
 u_0 K_0 = u K \quad \text{and} \quad \frac{1}{q^2}\frac{u_0}{K_0} = \frac{u}{K}\;,
\]
leading to
\[
 K=q K_0  \quad \text{and} \quad u=u_0/q \;.
\]
Therefore the Luttinger parameter $K$ of the clustered Luttinger
liquid is proportional to $q$ and the decay exponent of the
density-density correlation function is effectively suppressed by
$1/q$, see discussion below \eq{eq:nnLut}.


\begin{thebibliography}{39}%
\makeatletter
\providecommand \@ifxundefined [1]{%
 \@ifx{#1\undefined}
}%
\providecommand \@ifnum [1]{%
 \ifnum #1\expandafter \@firstoftwo
 \else \expandafter \@secondoftwo
 \fi
}%
\providecommand \@ifx [1]{%
 \ifx #1\expandafter \@firstoftwo
 \else \expandafter \@secondoftwo
 \fi
}%
\providecommand \natexlab [1]{#1}%
\providecommand \enquote  [1]{``#1''}%
\providecommand \bibnamefont  [1]{#1}%
\providecommand \bibfnamefont [1]{#1}%
\providecommand \citenamefont [1]{#1}%
\providecommand \href@noop [0]{\@secondoftwo}%
\providecommand \href [0]{\begingroup \@sanitize@url \@href}%
\providecommand \@href[1]{\@@startlink{#1}\@@href}%
\providecommand \@@href[1]{\endgroup#1\@@endlink}%
\providecommand \@sanitize@url [0]{\catcode `\\12\catcode `\$12\catcode
  `\&12\catcode `\#12\catcode `\^12\catcode `\_12\catcode `\%12\relax}%
\providecommand \@@startlink[1]{}%
\providecommand \@@endlink[0]{}%
\providecommand \url  [0]{\begingroup\@sanitize@url \@url }%
\providecommand \@url [1]{\endgroup\@href {#1}{\urlprefix }}%
\providecommand \urlprefix  [0]{URL }%
\providecommand \Eprint [0]{\href }%
\providecommand \doibase [0]{http://dx.doi.org/}%
\providecommand \selectlanguage [0]{\@gobble}%
\providecommand \bibinfo  [0]{\@secondoftwo}%
\providecommand \bibfield  [0]{\@secondoftwo}%
\providecommand \translation [1]{[#1]}%
\providecommand \BibitemOpen [0]{}%
\providecommand \bibitemStop [0]{}%
\providecommand \bibitemNoStop [0]{.\EOS\space}%
\providecommand \EOS [0]{\spacefactor3000\relax}%
\providecommand \BibitemShut  [1]{\csname bibitem#1\endcsname}%
\let\auto@bib@innerbib\@empty
\bibitem [{\citenamefont {Ng}\ and\ \citenamefont
  {Vanderbilt}(1995)}]{ng_stability_1995}%
  \BibitemOpen
  \bibfield  {author} {\bibinfo {author} {\bibfnamefont {K.~O.}\ \bibnamefont
  {Ng}}\ and\ \bibinfo {author} {\bibfnamefont {D.}~\bibnamefont
  {Vanderbilt}},\ }\href@noop {} {\bibfield  {journal} {\bibinfo  {journal}
  {Phys. Rev. B}\ }\textbf {\bibinfo {volume} {52}},\ \bibinfo {pages} {2177}
  (\bibinfo {year} {1995})}\BibitemShut {NoStop}%
\bibitem [{\citenamefont {Marchenko}(1986)}]{marchenko_1986}%
  \BibitemOpen
  \bibfield  {author} {\bibinfo {author} {\bibfnamefont {V.~I.}\ \bibnamefont
  {Marchenko}},\ }\href@noop {} {\bibfield  {journal} {\bibinfo  {journal} {Zh.
  Eksp. Teor. Fiz.}\ }\textbf {\bibinfo {volume} {90}},\ \bibinfo {pages}
  {2241} (\bibinfo {year} {1986})}\BibitemShut {NoStop}%
\bibitem [{\citenamefont
  {Rosenweig}(1985)}]{rosenweig_ferrohydrodynamics_1985}%
  \BibitemOpen
  \bibfield  {author} {\bibinfo {author} {\bibfnamefont {R.~E.}\ \bibnamefont
  {Rosenweig}},\ }\href@noop {} {\emph {\bibinfo {title}
  {Ferrohydrodynamics}}}\ (\bibinfo  {publisher} {Cambridge University Press},\
  \bibinfo {address} {Cambridge},\ \bibinfo {year} {1985})\BibitemShut
  {NoStop}%
\bibitem [{\citenamefont {Spivak}\ and\ \citenamefont
  {Kivelson}(2004)}]{spivak_phases_2004}%
  \BibitemOpen
  \bibfield  {author} {\bibinfo {author} {\bibfnamefont {B.}~\bibnamefont
  {Spivak}}\ and\ \bibinfo {author} {\bibfnamefont {S.~A.}\ \bibnamefont
  {Kivelson}},\ }\href@noop {} {\bibfield  {journal} {\bibinfo  {journal}
  {Phys. Rev. B}\ }\textbf {\bibinfo {volume} {70}},\ \bibinfo {pages} {155114}
  (\bibinfo {year} {2004})}\BibitemShut {NoStop}%
\bibitem [{\citenamefont {Spivak}\ \emph {et~al.}(2010)\citenamefont {Spivak},
  \citenamefont {Kravchenko}, \citenamefont {Kivelson},\ and\ \citenamefont
  {Gao}}]{spivak_colloquium_2010}%
  \BibitemOpen
  \bibfield  {author} {\bibinfo {author} {\bibfnamefont {B.}~\bibnamefont
  {Spivak}}, \bibinfo {author} {\bibfnamefont {S.~V.}\ \bibnamefont
  {Kravchenko}}, \bibinfo {author} {\bibfnamefont {S.~A.}\ \bibnamefont
  {Kivelson}}, \ and\ \bibinfo {author} {\bibfnamefont {X.~P.~A.}\ \bibnamefont
  {Gao}},\ }\href@noop {} {\bibfield  {journal} {\bibinfo  {journal} {Rev. Mod.
  Phys.}\ }\textbf {\bibinfo {volume} {82}},\ \bibinfo {pages} {1743} (\bibinfo
  {year} {2010})}\BibitemShut {NoStop}%
\bibitem [{\citenamefont {Koulakov}\ \emph {et~al.}(1996)\citenamefont
  {Koulakov}, \citenamefont {Fogler},\ and\ \citenamefont
  {Shklovskii}}]{koulakov_charge_1996}%
  \BibitemOpen
  \bibfield  {author} {\bibinfo {author} {\bibfnamefont {A.~A.}\ \bibnamefont
  {Koulakov}}, \bibinfo {author} {\bibfnamefont {M.~M.}\ \bibnamefont
  {Fogler}}, \ and\ \bibinfo {author} {\bibfnamefont {B.~I.}\ \bibnamefont
  {Shklovskii}},\ }\href@noop {} {\bibfield  {journal} {\bibinfo  {journal}
  {Phys. Rev. Lett.}\ }\textbf {\bibinfo {volume} {76}},\ \bibinfo {pages}
  {499} (\bibinfo {year} {1996})}\BibitemShut {NoStop}%
\bibitem [{\citenamefont {Moessner}\ and\ \citenamefont
  {Chalker}(1996)}]{moessner_exact_1996}%
  \BibitemOpen
  \bibfield  {author} {\bibinfo {author} {\bibfnamefont {R.}~\bibnamefont
  {Moessner}}\ and\ \bibinfo {author} {\bibfnamefont {J.~T.}\ \bibnamefont
  {Chalker}},\ }\href@noop {} {\bibfield  {journal} {\bibinfo  {journal} {Phys.
  Rev. B}\ }\textbf {\bibinfo {volume} {54}},\ \bibinfo {pages} {5006}
  (\bibinfo {year} {1996})}\BibitemShut {NoStop}%
\bibitem [{\citenamefont {Fogler}(2002)}]{fogler_stripe_2002}%
  \BibitemOpen
  \bibfield  {author} {\bibinfo {author} {\bibfnamefont {M.~M.}\ \bibnamefont
  {Fogler}},\ }in\ \href@noop {} {\emph {\bibinfo {booktitle} {High Magnetic
  Fields: Applications in Condensed Matter Physics and Spectroscopy}}}\
  (\bibinfo  {publisher} {Springer},\ \bibinfo {address} {Berlin},\ \bibinfo
  {year} {2002})\ \bibinfo {edition} {1st}\ ed.,\ pp.\ \bibinfo {pages}
  {98--138}\BibitemShut {NoStop}%
\bibitem [{\citenamefont {Eisenstein}\ \emph {et~al.}(2002)\citenamefont
  {Eisenstein}, \citenamefont {Cooper}, \citenamefont {Pfeiffer},\ and\
  \citenamefont {West}}]{eisenstein_insulating_2002}%
  \BibitemOpen
  \bibfield  {author} {\bibinfo {author} {\bibfnamefont {J.~P.}\ \bibnamefont
  {Eisenstein}}, \bibinfo {author} {\bibfnamefont {K.~B.}\ \bibnamefont
  {Cooper}}, \bibinfo {author} {\bibfnamefont {L.~N.}\ \bibnamefont
  {Pfeiffer}}, \ and\ \bibinfo {author} {\bibfnamefont {K.~W.}\ \bibnamefont
  {West}},\ }\href@noop {} {\bibfield  {journal} {\bibinfo  {journal} {Phys.
  Rev. Lett.}\ }\textbf {\bibinfo {volume} {88}},\ \bibinfo {pages} {076801}
  (\bibinfo {year} {2002})}\BibitemShut {NoStop}%
\bibitem [{\citenamefont {Santos}\ \emph {et~al.}(2000)\citenamefont {Santos},
  \citenamefont {Shlyapnikov}, \citenamefont {Zoller},\ and\ \citenamefont
  {Lewenstein}}]{santos_bose-einstein_2000}%
  \BibitemOpen
  \bibfield  {author} {\bibinfo {author} {\bibfnamefont {L.}~\bibnamefont
  {Santos}}, \bibinfo {author} {\bibfnamefont {G.~V.}\ \bibnamefont
  {Shlyapnikov}}, \bibinfo {author} {\bibfnamefont {P.}~\bibnamefont {Zoller}},
  \ and\ \bibinfo {author} {\bibfnamefont {M.}~\bibnamefont {Lewenstein}},\
  }\href@noop {} {\bibfield  {journal} {\bibinfo  {journal} {Phys. Rev. Lett.}\
  }\textbf {\bibinfo {volume} {85}},\ \bibinfo {pages} {1791} (\bibinfo {year}
  {2000})}\BibitemShut {NoStop}%
\bibitem [{\citenamefont {Baranov}\ \emph {et~al.}(2002)\citenamefont
  {Baranov}, \citenamefont {Mar’enko}, \citenamefont {Rychkov},\ and\
  \citenamefont {Shlyapnikov}}]{baranov_superfluid_2002}%
  \BibitemOpen
  \bibfield  {author} {\bibinfo {author} {\bibfnamefont {M.~A.}\ \bibnamefont
  {Baranov}}, \bibinfo {author} {\bibfnamefont {M.~S.}\ \bibnamefont
  {Mar’enko}}, \bibinfo {author} {\bibfnamefont {V.~S.}\ \bibnamefont
  {Rychkov}}, \ and\ \bibinfo {author} {\bibfnamefont {G.~V.}\ \bibnamefont
  {Shlyapnikov}},\ }\href@noop {} {\bibfield  {journal} {\bibinfo  {journal}
  {Phys. Rev. A}\ }\textbf {\bibinfo {volume} {66}},\ \bibinfo {pages} {013606}
  (\bibinfo {year} {2002})}\BibitemShut {NoStop}%
\bibitem [{\citenamefont {Doyle}\ \emph {et~al.}(2004)\citenamefont {Doyle},
  \citenamefont {Friedrich}, \citenamefont {Krems},\ and\ \citenamefont
  {{Masnou-Seeuws}}}]{doyle_editorial_2004}%
  \BibitemOpen
  \bibfield  {author} {\bibinfo {author} {\bibfnamefont {J.}~\bibnamefont
  {Doyle}}, \bibinfo {author} {\bibfnamefont {B.}~\bibnamefont {Friedrich}},
  \bibinfo {author} {\bibfnamefont {R.~V.}\ \bibnamefont {Krems}}, \ and\
  \bibinfo {author} {\bibfnamefont {F.}~\bibnamefont {{Masnou-Seeuws}}},\
  }\href@noop {} {\bibfield  {journal} {\bibinfo  {journal} {Eur. Phys. J. D}\
  }\textbf {\bibinfo {volume} {31}},\ \bibinfo {pages} {149} (\bibinfo {year}
  {2004})}\BibitemShut {NoStop}%
\bibitem [{\citenamefont {Baranov}(2008)}]{baranov_theoretical_2008}%
  \BibitemOpen
  \bibfield  {author} {\bibinfo {author} {\bibfnamefont {M.}~\bibnamefont
  {Baranov}},\ }\href@noop {} {\bibfield  {journal} {\bibinfo  {journal} {Phys.
  Rep.}\ }\textbf {\bibinfo {volume} {464}},\ \bibinfo {pages} {71} (\bibinfo
  {year} {2008})}\BibitemShut {NoStop}%
\bibitem [{\citenamefont {Carr}\ \emph {et~al.}(2009)\citenamefont {Carr},
  \citenamefont {{DeMille}}, \citenamefont {Krems},\ and\ \citenamefont
  {Ye}}]{carr_cold_2009}%
  \BibitemOpen
  \bibfield  {author} {\bibinfo {author} {\bibfnamefont {L.~D.}\ \bibnamefont
  {Carr}}, \bibinfo {author} {\bibfnamefont {D.}~\bibnamefont {{DeMille}}},
  \bibinfo {author} {\bibfnamefont {R.~V.}\ \bibnamefont {Krems}}, \ and\
  \bibinfo {author} {\bibfnamefont {J.}~\bibnamefont {Ye}},\ }\href@noop {}
  {\bibfield  {journal} {\bibinfo  {journal} {New J. Phys.}\ }\textbf {\bibinfo
  {volume} {11}},\ \bibinfo {pages} {055049} (\bibinfo {year}
  {2009})}\BibitemShut {NoStop}%
\bibitem [{\citenamefont {Lahaye}\ \emph {et~al.}(2009)\citenamefont {Lahaye},
  \citenamefont {Menotti}, \citenamefont {Santos}, \citenamefont {Lewenstein},\
  and\ \citenamefont {Pfau}}]{lahaye_physics_2009}%
  \BibitemOpen
  \bibfield  {author} {\bibinfo {author} {\bibfnamefont {T.}~\bibnamefont
  {Lahaye}}, \bibinfo {author} {\bibfnamefont {C.}~\bibnamefont {Menotti}},
  \bibinfo {author} {\bibfnamefont {L.}~\bibnamefont {Santos}}, \bibinfo
  {author} {\bibfnamefont {M.}~\bibnamefont {Lewenstein}}, \ and\ \bibinfo
  {author} {\bibfnamefont {T.}~\bibnamefont {Pfau}},\ }\href@noop {} {\bibfield
   {journal} {\bibinfo  {journal} {Rep. Prog. Phys.}\ }\textbf {\bibinfo
  {volume} {72}},\ \bibinfo {pages} {126401} (\bibinfo {year}
  {2009})}\BibitemShut {NoStop}%
\bibitem [{\citenamefont {Wunsch}\ \emph {et~al.}(2011)\citenamefont {Wunsch},
  \citenamefont {Zinner}, \citenamefont {Mekhov}, \citenamefont {Huang},
  \citenamefont {Wang},\ and\ \citenamefont {Demler}}]{wunsch_few-body_2011}%
  \BibitemOpen
  \bibfield  {author} {\bibinfo {author} {\bibfnamefont {B.}~\bibnamefont
  {Wunsch}}, \bibinfo {author} {\bibfnamefont {N.~T.}\ \bibnamefont {Zinner}},
  \bibinfo {author} {\bibfnamefont {I.~B.}\ \bibnamefont {Mekhov}}, \bibinfo
  {author} {\bibfnamefont {S.~J.}\ \bibnamefont {Huang}}, \bibinfo {author}
  {\bibfnamefont {D.~W.}\ \bibnamefont {Wang}}, \ and\ \bibinfo {author}
  {\bibfnamefont {E.}~\bibnamefont {Demler}},\ }\href@noop {} {\bibfield
  {journal} {\bibinfo  {journal} {Phys. Rev. Lett.}\ }\textbf {\bibinfo
  {volume} {107}},\ \bibinfo {pages} {073201} (\bibinfo {year}
  {2011})}\BibitemShut {NoStop}%
\bibitem [{\citenamefont {Dalmonte}\ \emph {et~al.}(2011)\citenamefont
  {Dalmonte}, \citenamefont {Zoller},\ and\ \citenamefont
  {Pupillo}}]{dalmonte_trimer_2011}%
  \BibitemOpen
  \bibfield  {author} {\bibinfo {author} {\bibfnamefont {M.}~\bibnamefont
  {Dalmonte}}, \bibinfo {author} {\bibfnamefont {P.}~\bibnamefont {Zoller}}, \
  and\ \bibinfo {author} {\bibfnamefont {G.}~\bibnamefont {Pupillo}},\
  }\href@noop {} {\bibfield  {journal} {\bibinfo  {journal} {Phys. Rev. Lett.}\
  }\textbf {\bibinfo {volume} {107}},\ \bibinfo {pages} {163202} (\bibinfo
  {year} {2011})}\BibitemShut {NoStop}%
\bibitem [{\citenamefont {Bauer}\ and\ \citenamefont
  {Parish}(2012)}]{bauer_dipolar_2012}%
  \BibitemOpen
  \bibfield  {author} {\bibinfo {author} {\bibfnamefont {M.}~\bibnamefont
  {Bauer}}\ and\ \bibinfo {author} {\bibfnamefont {M.~M.}\ \bibnamefont
  {Parish}},\ }\href@noop {} {\bibfield  {journal} {\bibinfo  {journal} {Phys.
  Rev. Lett.}\ }\textbf {\bibinfo {volume} {108}},\ \bibinfo {pages} {255302}
  (\bibinfo {year} {2012})}\BibitemShut {NoStop}%
\bibitem [{\citenamefont {Chotia}\ \emph {et~al.}(2012)\citenamefont {Chotia},
  \citenamefont {Neyenhuis}, \citenamefont {Moses}, \citenamefont {Yan},
  \citenamefont {Covey}, \citenamefont {{Foss-Feig}}, \citenamefont {Rey},
  \citenamefont {Jin},\ and\ \citenamefont {Ye}}]{chotia_long-lived_2012}%
  \BibitemOpen
  \bibfield  {author} {\bibinfo {author} {\bibfnamefont {A.}~\bibnamefont
  {Chotia}}, \bibinfo {author} {\bibfnamefont {B.}~\bibnamefont {Neyenhuis}},
  \bibinfo {author} {\bibfnamefont {S.~A.}\ \bibnamefont {Moses}}, \bibinfo
  {author} {\bibfnamefont {B.}~\bibnamefont {Yan}}, \bibinfo {author}
  {\bibfnamefont {J.~P.}\ \bibnamefont {Covey}}, \bibinfo {author}
  {\bibfnamefont {M.}~\bibnamefont {{Foss-Feig}}}, \bibinfo {author}
  {\bibfnamefont {A.~M.}\ \bibnamefont {Rey}}, \bibinfo {author} {\bibfnamefont
  {D.~S.}\ \bibnamefont {Jin}}, \ and\ \bibinfo {author} {\bibfnamefont
  {J.}~\bibnamefont {Ye}},\ }\href@noop {} {\bibfield  {journal} {\bibinfo
  {journal} {Phys. Rev. Lett.}\ }\textbf {\bibinfo {volume} {108}},\ \bibinfo
  {pages} {080405} (\bibinfo {year} {2012})}\BibitemShut {NoStop}%
\bibitem [{sup()}]{supp}%
  \BibitemOpen
  \href@noop {} {}\bibinfo {note} {See supplementary material}\BibitemShut
  {NoStop}%
\bibitem [{\citenamefont {Giamarchi}(2004)}]{giamarchi_quantum_2004}%
  \BibitemOpen
  \bibfield  {author} {\bibinfo {author} {\bibfnamefont {T.}~\bibnamefont
  {Giamarchi}},\ }\href@noop {} {\emph {\bibinfo {title} {Quantum Physics in
  One Dimension}}}\ (\bibinfo  {publisher} {Oxford University Press, {USA}},\
  \bibinfo {year} {2004})\BibitemShut {NoStop}%
\bibitem [{\citenamefont {White}(1992)}]{white_density_1992}%
  \BibitemOpen
  \bibfield  {author} {\bibinfo {author} {\bibfnamefont {S.~R.}\ \bibnamefont
  {White}},\ }\href@noop {} {\bibfield  {journal} {\bibinfo  {journal} {Phys.
  Rev. Lett.}\ }\textbf {\bibinfo {volume} {69}},\ \bibinfo {pages} {2863}
  (\bibinfo {year} {1992})}\BibitemShut {NoStop}%
\bibitem [{\citenamefont
  {Schollw{\"o}ck}(2005)}]{schollwoeck_density-matrix_2005}%
  \BibitemOpen
  \bibfield  {author} {\bibinfo {author} {\bibfnamefont {U.}~\bibnamefont
  {Schollw{\"o}ck}},\ }\href@noop {} {\bibfield  {journal} {\bibinfo  {journal}
  {Rev. Mod. Phys.}\ }\textbf {\bibinfo {volume} {77}},\ \bibinfo {pages} {259}
  (\bibinfo {year} {2005})}\BibitemShut {NoStop}%
\bibitem [{\citenamefont {M{\"u}ller}\ \emph {et~al.}(2001)\citenamefont
  {M{\"u}ller}, \citenamefont {Jonckheere}, \citenamefont {Miniatura},\ and\
  \citenamefont {Delande}}]{mueller_weak_2001}%
  \BibitemOpen
  \bibfield  {author} {\bibinfo {author} {\bibfnamefont {C.~A.}\ \bibnamefont
  {M{\"u}ller}}, \bibinfo {author} {\bibfnamefont {T.}~\bibnamefont
  {Jonckheere}}, \bibinfo {author} {\bibfnamefont {C.}~\bibnamefont
  {Miniatura}}, \ and\ \bibinfo {author} {\bibfnamefont {D.}~\bibnamefont
  {Delande}},\ }\href@noop {} {\bibfield  {journal} {\bibinfo  {journal} {Phys.
  Rev. A}\ }\textbf {\bibinfo {volume} {64}},\ \bibinfo {pages} {053804}
  (\bibinfo {year} {2001})}\BibitemShut {NoStop}%
\bibitem [{\citenamefont {Altman}\ \emph {et~al.}(2004)\citenamefont {Altman},
  \citenamefont {Demler},\ and\ \citenamefont {Lukin}}]{altman_probing_2004}%
  \BibitemOpen
  \bibfield  {author} {\bibinfo {author} {\bibfnamefont {E.}~\bibnamefont
  {Altman}}, \bibinfo {author} {\bibfnamefont {E.}~\bibnamefont {Demler}}, \
  and\ \bibinfo {author} {\bibfnamefont {M.~D.}\ \bibnamefont {Lukin}},\
  }\href@noop {} {\bibfield  {journal} {\bibinfo  {journal} {Phys. Rev. A}\
  }\textbf {\bibinfo {volume} {70}},\ \bibinfo {pages} {013603} (\bibinfo
  {year} {2004})}\BibitemShut {NoStop}%
\bibitem [{\citenamefont {Zinner}\ \emph {et~al.}(2011)\citenamefont {Zinner},
  \citenamefont {Wunsch}, \citenamefont {Mekhov}, \citenamefont {Huang},
  \citenamefont {Wang},\ and\ \citenamefont {Demler}}]{zinner_few-body_2011}%
  \BibitemOpen
  \bibfield  {author} {\bibinfo {author} {\bibfnamefont {N.~T.}\ \bibnamefont
  {Zinner}}, \bibinfo {author} {\bibfnamefont {B.}~\bibnamefont {Wunsch}},
  \bibinfo {author} {\bibfnamefont {I.~B.}\ \bibnamefont {Mekhov}}, \bibinfo
  {author} {\bibfnamefont {S.~J.}~\bibnamefont {Huang}}, \bibinfo {author}
  {\bibfnamefont {D.~W.}~\bibnamefont {Wang}}, \ and\ \bibinfo {author}
  {\bibfnamefont {E.}~\bibnamefont {Demler}},\ }\href@noop {} {\bibfield
  {journal} {\bibinfo  {journal} {Phys. Rev. A}\ }\textbf {\bibinfo {volume}
  {84}},\ \bibinfo {pages} {063606} (\bibinfo {year} {2011})}\BibitemShut
  {NoStop}%
\bibitem [{\citenamefont {St{\"o}ferle}\ \emph {et~al.}(2004)\citenamefont
  {St{\"o}ferle}, \citenamefont {Moritz}, \citenamefont {Schori}, \citenamefont
  {K{\"o}hl},\ and\ \citenamefont {Esslinger}}]{stoeferle_transition_2004}%
  \BibitemOpen
  \bibfield  {author} {\bibinfo {author} {\bibfnamefont {T.}~\bibnamefont
  {St{\"o}ferle}}, \bibinfo {author} {\bibfnamefont {H.}~\bibnamefont
  {Moritz}}, \bibinfo {author} {\bibfnamefont {C.}~\bibnamefont {Schori}},
  \bibinfo {author} {\bibfnamefont {M.}~\bibnamefont {K{\"o}hl}}, \ and\
  \bibinfo {author} {\bibfnamefont {T.}~\bibnamefont {Esslinger}},\ }\href@noop
  {} {\bibfield  {journal} {\bibinfo  {journal} {Phys. Rev. Lett.}\ }\textbf
  {\bibinfo {volume} {92}},\ \bibinfo {pages} {130403} (\bibinfo {year}
  {2004})}\BibitemShut {NoStop}%
\bibitem [{\citenamefont {Regal}\ and\ \citenamefont
  {Jin}(2003)}]{regal_measurement_2003}%
  \BibitemOpen
  \bibfield  {author} {\bibinfo {author} {\bibfnamefont {C.~A.}\ \bibnamefont
  {Regal}}\ and\ \bibinfo {author} {\bibfnamefont {D.~S.}\ \bibnamefont
  {Jin}},\ }\href@noop {} {\bibfield  {journal} {\bibinfo  {journal} {Phys.
  Rev. Lett.}\ }\textbf {\bibinfo {volume} {90}},\ \bibinfo {pages} {230404}
  (\bibinfo {year} {2003})}\BibitemShut {NoStop}%
\bibitem [{\citenamefont {Gupta}\ \emph {et~al.}(2003)\citenamefont {Gupta},
  \citenamefont {Hadzibabic}, \citenamefont {Zwierlein}, \citenamefont {Stan},
  \citenamefont {Dieckmann}, \citenamefont {Schunck}, \citenamefont {van
  Kempen}, \citenamefont {Verhaar},\ and\ \citenamefont
  {Ketterle}}]{gupta_radio-frequency_2003}%
  \BibitemOpen
  \bibfield  {author} {\bibinfo {author} {\bibfnamefont {S.}~\bibnamefont
  {Gupta}}, \bibinfo {author} {\bibfnamefont {Z.}~\bibnamefont {Hadzibabic}},
  \bibinfo {author} {\bibfnamefont {M.~W.}\ \bibnamefont {Zwierlein}}, \bibinfo
  {author} {\bibfnamefont {C.~A.}\ \bibnamefont {Stan}}, \bibinfo {author}
  {\bibfnamefont {K.}~\bibnamefont {Dieckmann}}, \bibinfo {author}
  {\bibfnamefont {C.~H.}\ \bibnamefont {Schunck}}, \bibinfo {author}
  {\bibfnamefont {E.~G.~M.}\ \bibnamefont {van Kempen}}, \bibinfo {author}
  {\bibfnamefont {B.~J.}\ \bibnamefont {Verhaar}}, \ and\ \bibinfo {author}
  {\bibfnamefont {W.}~\bibnamefont {Ketterle}},\ }\href@noop {} {\bibfield
  {journal} {\bibinfo  {journal} {Science}\ }\textbf {\bibinfo {volume}
  {300}},\ \bibinfo {pages} {1723 } (\bibinfo {year} {2003})}\BibitemShut
  {NoStop}%
\bibitem [{\citenamefont {Regal}\ \emph {et~al.}(2003)\citenamefont {Regal},
  \citenamefont {Ticknor}, \citenamefont {Bohn},\ and\ \citenamefont
  {Jin}}]{regal_creation_2003}%
  \BibitemOpen
  \bibfield  {author} {\bibinfo {author} {\bibfnamefont {C.~A.}\ \bibnamefont
  {Regal}}, \bibinfo {author} {\bibfnamefont {C.}~\bibnamefont {Ticknor}},
  \bibinfo {author} {\bibfnamefont {J.~L.}\ \bibnamefont {Bohn}}, \ and\
  \bibinfo {author} {\bibfnamefont {D.~S.}\ \bibnamefont {Jin}},\ }\href@noop
  {} {\bibfield  {journal} {\bibinfo  {journal} {Nature (London)}\ }\textbf
  {\bibinfo {volume} {424}},\ \bibinfo {pages} {47} (\bibinfo {year}
  {2003})}\BibitemShut {NoStop}%
\bibitem [{\citenamefont {Ewald}(1921)}]{ewald_berechnung_1921}%
  \BibitemOpen
  \bibfield  {author} {\bibinfo {author} {\bibfnamefont {P.~P.}\ \bibnamefont
  {Ewald}},\ }\href@noop {} {\bibfield  {journal} {\bibinfo  {journal} {Ann.
  Phys.}\ }\textbf {\bibinfo {volume} {369}},\ \bibinfo {pages} {253} (\bibinfo
  {year} {1921})}\BibitemShut {NoStop}%
\bibitem [{\citenamefont {Essmann}\ \emph {et~al.}(1995)\citenamefont
  {Essmann}, \citenamefont {Perera}, \citenamefont {Berkowitz}, \citenamefont
  {Darden}, \citenamefont {Lee},\ and\ \citenamefont
  {Pedersen}}]{essmann_smooth_1995}%
  \BibitemOpen
  \bibfield  {author} {\bibinfo {author} {\bibfnamefont {U.}~\bibnamefont
  {Essmann}}, \bibinfo {author} {\bibfnamefont {L.}~\bibnamefont {Perera}},
  \bibinfo {author} {\bibfnamefont {M.~L.}\ \bibnamefont {Berkowitz}}, \bibinfo
  {author} {\bibfnamefont {T.}~\bibnamefont {Darden}}, \bibinfo {author}
  {\bibfnamefont {H.}~\bibnamefont {Lee}}, \ and\ \bibinfo {author}
  {\bibfnamefont {L.~G.}\ \bibnamefont {Pedersen}},\ }\href@noop {} {\bibfield
  {journal} {\bibinfo  {journal} {J. Chem. Phys.}\ }\textbf {\bibinfo {volume}
  {103}},\ \bibinfo {pages} {8577} (\bibinfo {year} {1995})}\BibitemShut
  {NoStop}%
\bibitem [{\citenamefont {Grzybowski}\ \emph {et~al.}(2000)\citenamefont
  {Grzybowski}, \citenamefont {Gw{\'o}{\'z}d{\'z}},\ and\ \citenamefont
  {Br{\'o}dka}}]{grzybowski_ewald_2000}%
  \BibitemOpen
  \bibfield  {author} {\bibinfo {author} {\bibfnamefont {A.}~\bibnamefont
  {Grzybowski}}, \bibinfo {author} {\bibfnamefont {E.}~\bibnamefont
  {Gw{\'o}{\'z}d{\'z}}}, \ and\ \bibinfo {author} {\bibfnamefont
  {A.}~\bibnamefont {Br{\'o}dka}},\ }\href@noop {} {\bibfield  {journal}
  {\bibinfo  {journal} {Phys. Rev. B}\ }\textbf {\bibinfo {volume} {61}},\
  \bibinfo {pages} {6706} (\bibinfo {year} {2000})}\BibitemShut {NoStop}%
\bibitem [{\citenamefont {Ni}\ \emph {et~al.}(2008)\citenamefont {Ni},
  \citenamefont {Ospelkaus}, \citenamefont {de~Miranda}, \citenamefont {Pe'er},
  \citenamefont {Neyenhuis}, \citenamefont {Zirbel}, \citenamefont
  {Kotochigova}, \citenamefont {Julienne}, \citenamefont {Jin},\ and\
  \citenamefont {Ye}}]{ni_high_2008}%
  \BibitemOpen
  \bibfield  {author} {\bibinfo {author} {\bibfnamefont {K.}~\bibnamefont
  {Ni}}, \bibinfo {author} {\bibfnamefont {S.}~\bibnamefont {Ospelkaus}},
  \bibinfo {author} {\bibfnamefont {M.~H.~G.}\ \bibnamefont {de~Miranda}},
  \bibinfo {author} {\bibfnamefont {A.}~\bibnamefont {Pe'er}}, \bibinfo
  {author} {\bibfnamefont {B.}~\bibnamefont {Neyenhuis}}, \bibinfo {author}
  {\bibfnamefont {J.~J.}\ \bibnamefont {Zirbel}}, \bibinfo {author}
  {\bibfnamefont {S.}~\bibnamefont {Kotochigova}}, \bibinfo {author}
  {\bibfnamefont {P.~S.}\ \bibnamefont {Julienne}}, \bibinfo {author}
  {\bibfnamefont {D.~S.}\ \bibnamefont {Jin}}, \ and\ \bibinfo {author}
  {\bibfnamefont {J.}~\bibnamefont {Ye}},\ }\href@noop {} {\bibfield  {journal}
  {\bibinfo  {journal} {Science}\ }\textbf {\bibinfo {volume} {322}},\ \bibinfo
  {pages} {231 } (\bibinfo {year} {2008})}\BibitemShut {NoStop}%
\bibitem [{\citenamefont {Debatin}\ \emph {et~al.}(2011)\citenamefont
  {Debatin}, \citenamefont {Takekoshi}, \citenamefont {Rameshan}, \citenamefont
  {Reichs{\"o}llner}, \citenamefont {Ferlaino}, \citenamefont {Grimm},
  \citenamefont {Vexiau}, \citenamefont {Bouloufa}, \citenamefont {Dulieu},\
  and\ \citenamefont {N{\"a}gerl}}]{debatin_molecular_2011}%
  \BibitemOpen
  \bibfield  {author} {\bibinfo {author} {\bibfnamefont {M.}~\bibnamefont
  {Debatin}}, \bibinfo {author} {\bibfnamefont {T.}~\bibnamefont {Takekoshi}},
  \bibinfo {author} {\bibfnamefont {R.}~\bibnamefont {Rameshan}}, \bibinfo
  {author} {\bibfnamefont {L.}~\bibnamefont {Reichs{\"o}llner}}, \bibinfo
  {author} {\bibfnamefont {F.}~\bibnamefont {Ferlaino}}, \bibinfo {author}
  {\bibfnamefont {R.}~\bibnamefont {Grimm}}, \bibinfo {author} {\bibfnamefont
  {R.}~\bibnamefont {Vexiau}}, \bibinfo {author} {\bibfnamefont
  {N.}~\bibnamefont {Bouloufa}}, \bibinfo {author} {\bibfnamefont
  {O.}~\bibnamefont {Dulieu}}, \ and\ \bibinfo {author} {\bibfnamefont
  {H.}~\bibnamefont {N{\"a}gerl}},\ }\href@noop {} {\bibfield  {journal}
  {\bibinfo  {journal} {Phys. Chem. Chem. Phys.}\ }\textbf {\bibinfo {volume}
  {13}},\ \bibinfo {pages} {18926} (\bibinfo {year} {2011})}\BibitemShut
  {NoStop}%
\bibitem [{\citenamefont {Park}\ \emph {et~al.}(2012)\citenamefont {Park},
  \citenamefont {Wu}, \citenamefont {Santiago}, \citenamefont {Tiecke},
  \citenamefont {Ahmadi},\ and\ \citenamefont {Zwierlein}}]{park_quantum_2011}%
  \BibitemOpen
  \bibfield  {author} {\bibinfo {author} {\bibfnamefont {J.~W.}\ \bibnamefont
  {Park}}, \bibinfo {author} {\bibfnamefont {C.}~\bibnamefont {Wu}}, \bibinfo
  {author} {\bibfnamefont {I.}~\bibnamefont {Santiago}}, \bibinfo {author}
  {\bibfnamefont {T.~G.}\ \bibnamefont {Tiecke}}, \bibinfo {author}
  {\bibfnamefont {P.}~\bibnamefont {Ahmadi}}, \ and\ \bibinfo {author}
  {\bibfnamefont {M.~W.}\ \bibnamefont {Zwierlein}},\ }\href@noop {} {\bibfield
   {journal} {\bibinfo  {journal} {Phys. Rev. A}\ }\textbf {\bibinfo {volume}
  {85}},\ \bibinfo {pages} {051602(R)} (\bibinfo {year} {2012})}\BibitemShut
  {NoStop}%
\bibitem [{\citenamefont {Haimberger}\ \emph {et~al.}(2006)\citenamefont
  {Haimberger}, \citenamefont {Kleinert}, \citenamefont {Dulieu},\ and\
  \citenamefont {Bigelow}}]{haimberger_processes_2006}%
  \BibitemOpen
  \bibfield  {author} {\bibinfo {author} {\bibfnamefont {C.}~\bibnamefont
  {Haimberger}}, \bibinfo {author} {\bibfnamefont {J.}~\bibnamefont
  {Kleinert}}, \bibinfo {author} {\bibfnamefont {O.}~\bibnamefont {Dulieu}}, \
  and\ \bibinfo {author} {\bibfnamefont {N.~P.}\ \bibnamefont {Bigelow}},\
  }\href@noop {} {\bibfield  {journal} {\bibinfo  {journal} {J. Phys. B: At.
  Mol. Opt. Phys.}\ }\textbf {\bibinfo {volume} {39}},\ \bibinfo {pages} {S957}
  (\bibinfo {year} {2006})}\BibitemShut {NoStop}%
\bibitem [{\citenamefont {Deiglmayr}\ \emph {et~al.}(2009)\citenamefont
  {Deiglmayr}, \citenamefont {Repp}, \citenamefont {Grochola}, \citenamefont
  {M{\"o}rtlbauer}, \citenamefont {Gl{\"u}ck}, \citenamefont {Dulieu},
  \citenamefont {Lange}, \citenamefont {Wester},\ and\ \citenamefont
  {Weidem{\"u}ller}}]{deiglmayr_formation_2009}%
  \BibitemOpen
  \bibfield  {author} {\bibinfo {author} {\bibfnamefont {J.}~\bibnamefont
  {Deiglmayr}}, \bibinfo {author} {\bibfnamefont {M.}~\bibnamefont {Repp}},
  \bibinfo {author} {\bibfnamefont {A.}~\bibnamefont {Grochola}}, \bibinfo
  {author} {\bibfnamefont {K.}~\bibnamefont {M{\"o}rtlbauer}}, \bibinfo
  {author} {\bibfnamefont {C.}~\bibnamefont {Gl{\"u}ck}}, \bibinfo {author}
  {\bibfnamefont {O.}~\bibnamefont {Dulieu}}, \bibinfo {author} {\bibfnamefont
  {J.}~\bibnamefont {Lange}}, \bibinfo {author} {\bibfnamefont
  {R.}~\bibnamefont {Wester}}, \ and\ \bibinfo {author} {\bibfnamefont
  {M.}~\bibnamefont {Weidem{\"u}ller}},\ }\href@noop {} {\bibfield  {journal}
  {\bibinfo  {journal} {Faraday Discuss.}\ }\textbf {\bibinfo {volume} {142}},\
  \bibinfo {pages} {335} (\bibinfo {year} {2009})}\BibitemShut {NoStop}%
\bibitem [{\citenamefont {Kollath}\ \emph {et~al.}(2008)\citenamefont
  {Kollath}, \citenamefont {Meyer},\ and\ \citenamefont
  {Giamarchi}}]{kollath_dipolar_2008}%
  \BibitemOpen
  \bibfield  {author} {\bibinfo {author} {\bibfnamefont {C.}~\bibnamefont
  {Kollath}}, \bibinfo {author} {\bibfnamefont {J.~S.}\ \bibnamefont {Meyer}},
  \ and\ \bibinfo {author} {\bibfnamefont {T.}~\bibnamefont {Giamarchi}},\
  }\href@noop {} {\bibfield  {journal} {\bibinfo  {journal} {Phys. Rev. Lett.}\
  }\textbf {\bibinfo {volume} {100}},\ \bibinfo {pages} {130403} (\bibinfo
  {year} {2008})}\BibitemShut {NoStop}%
\end{thebibliography}%

\end{document}